\documentclass[aps,prx,reprint,superscriptaddress,amsmath,amssymb,longbibliography]{revtex4-1}

\usepackage{graphicx}
\usepackage{verbatim}
\usepackage{dcolumn}
\usepackage{bm}
\usepackage{xcolor}
\usepackage{hyperref}
\hypersetup{
    colorlinks,
    linkcolor={blue!50!black},
    citecolor={blue!50!black},
    urlcolor={blue!80!black}
}
\usepackage{mathtools}
\usepackage[shortlabels]{enumitem}
\usepackage{float}
\newtheorem{theorem}{Theorem}
\usepackage{dsfont}
\DeclarePairedDelimiter\bra{\langle}{\rvert}
\DeclarePairedDelimiter\ket{\lvert}{\rangle}

\newcommand{\id}{\mathds{1}}

\newcommand{\padA}{\affiliation{Dipartimento di Fisica e Astronomia “G. Galilei” \& Padua Quantum Technologies Research Center, Università degli Studi di Padova, Italy I-35131, Padova, Italy}}
\newcommand{\padB}{\affiliation{INFN, Sezione di Padova, via Marzolo 8, I-35131, Padova}}
\newcommand{\padC}{\affiliation{Institute for Complex Quantum Systems, Ulm University, Albert-Einstein-Allee 11, 89069 Ulm, Germany}}
\newcommand{\planqc}{\affiliation{PlanQC GmbH, Lichtenbergstr. 8, 85748 Garching, Germany}}

\usepackage{orcidlink}
\newcommand{\orciddavide}{\orcidlink{0000-0001-8219-5806}}
\newcommand{\orciddaniel}{\orcidlink{0000-0001-7658-3546}}
\newcommand{\orcidilaria}{\orcidlink{0000-0002-3806-2034}}
\newcommand{\orcidsimone}{\orcidlink{0000-0002-8882-2169}}
\newcommand{\orcidantonio}{\orcidlink{0000-0002-1596-9477}}

\begin{document}

\title{Quantum algorithms for compact polymer thermodynamics}

\author{Davide Rattacaso\orciddavide}\padA \padB
\author{Daniel Jaschke \orciddaniel}\padA \padB \planqc \padC 
\author{Antonio Trovato\orcidantonio}\padA \padB
\author{Ilaria Siloi\orcidilaria}\padA \padB
\author{Simone Montangero\orcidsimone}\padA \padB

\date{\today}

\begin{abstract}

Efficient sampling from ensembles of Hamiltonian cycles is critical for predicting the thermodynamic properties of compact polymers, with applications including modeling protein and RNA folding and designing soft materials. Although classical Monte Carlo methods are widely regarded as the standard approach, their efficiency is strongly limited when applied to compact polymers. In this work, we enable a quadratic speedup in the estimation of thermodynamic properties of maximally compact polymers and heteropolymers by quantum computation. To this end, we encode the target thermodynamic ensemble into the amplitudes of a quantum state, i.e., a quantum sample, which can be processed via amplitude amplification. Using quantum equational reasoning, we construct a local parent Hamiltonian whose unique ground state realizes a quantum sample of all Hamiltonian cycles. This state can be prepared on quantum hardware using ground-state preparation methods, such as quantum annealing, and subsequently manipulated to generate quantum samples of polymers and heteropolymers at a target temperature. Finally, we approximate the quantum sample as a tensor network, revealing an entanglement area law. For fixed-width rectangular lattices, we obtain a time-efficient and compact encoding of the full ensemble of Hamiltonian cycles, enabling the efficient evaluation of expectation values, partition functions, and configuration probabilities via tensor contractions, without resorting to sampling.
\end{abstract}

\maketitle

\section{Introduction}\label{sec:intro}

Hamiltonian paths are curves on a lattice that visit each vertex exactly once. These paths serve as mathematical models for the spatial organization of compact polymers~\cite{Vanderzande_1998,Nagle_74,Gorodn_1976,LUA2004717,PhysRevLett.112.118302}, such as globular proteins~\cite{PhysRevLett.100.118102} and packaged viral RNA~\cite{DYKEMAN2011399,Twarock2018,TWAROCK201874}. Understanding the statistical properties of the complete ensemble of such curves on a fixed lattice is essential for predicting their thermodynamic behavior.

Classical Monte Carlo (MC) methods are commonly employed to investigate polymer thermodynamics: sampling compact polymers poses significant challenges for MC~\cite{10.1063/1.470277,LUA2004717, PhysRevE.104.054503, Mansfield_sampling}. Transfer matrix methods~\cite{stoyan1996enumeration, Jacobsen_2007} provide an alternative to Monte Carlo sampling, but their applicability is restricted by the exponential increase in computational cost with the shorter dimensions of the two|or three|dimensional lattice

In recent years, quantum computation has emerged as a promising approach for efficiently simulating and predicting the physics of soft-matter systems described by lattice-path configurations~\cite{PhysRevLett.127.080501,doi:10.1126/sciadv.adi0204, PhysRevLett.134.158101}.
The standard approach is to encode the spatial configurations of polymers in the degenerate ground states of a classical spin Hamiltonian~\cite{PhysRevLett.127.080501,doi:10.1126/sciadv.adi0204},  which can be prepared by quantum annealing~\cite{Kadowaki_PRE1998,Farhi_SCI01,Santoro_SCI02,aqc_review}. However, this encoding requires an exponentially increasing number of spins and pairwise interactions when applied to modeling compact polymers. This exponential overhead is intrinsic to the topological nature of the problem: the existence of one single Hamiltonian path connecting all lattice sites cannot be enforced by a collection of purely local Boolean constraints~\cite{libkin2004elements}, that is, a collection of local classical spin interactions. As shown in Ref.~\cite{PhysRevLett.134.158101}, a possible solution is to construct a lattice gauge model with compact polymer configurations as degenerate ground states.


In addition to the potential acceleration in preparing classical configurations via quantum processes such as quantum annealing, quantum computation provides a provable quadratic speedup for sampling expectation values and partition functions of statistical ensembles using amplitude amplification~\cite{Brassard}. Achieving this speedup requires encoding the target probability distribution $p(\bm\sigma)$ over the space of classical configuration $\bm \sigma$ into the amplitudes of a corresponding quantum state $\ket{\psi}=\sum_{\bm\sigma} \sqrt{p(\bm\sigma)}\ket{\bm\sigma}$, referred to as a coherent quantum sample~\cite{doi:10.1098/rspa.2015.0301}. In this paper, we unlock this quantum advantage by departing from the standard quantum optimization paradigm, in which polymer configurations are encoded in the degenerate ground states of a classical Hamiltonian. Instead, we leverage equational reasoning~\cite{quantum_eq_reasoning} to construct a non-diagonal parent Hamiltonian $\hat{H}_\text{HC}$ whose unique zero-energy ground state, $\ket{X_\text{HC}}$, is a coherent quantum sample of all compact polymer configurations. Remarkably, the parent Hamiltonian that we construct is local, meaning that both the number of interactions per monomer and their range do not scale with system size.

Numerous algorithms could be employed to prepare the state $\ket{X_\text{HC}}$. Examples include quantum annealing~\cite{Kadowaki_PRE1998, Farhi_SCI01, Santoro_SCI02,aqc_review}, optimal control techniques~\cite{opt_c_0,opt_c_3,opt_c_4}, quantum approximate optimization algorithms~\cite{farhi2014quantum,farhi2019quantum,BLEKOS20241}, and dissipative time-evolution methods~\cite{mcardle2019variational, Motta2020, PRXQuantum.3.010320,gluza2025doublebracketquantumalgorithmsquantum,ding2025endtoendefficientquantumthermal}. The performance of these approaches depends on both the specific problem instance and the details of their implementation. 

The construction the parent Hamiltonian $\hat H_\text{HC}$ for $\ket{X_\text{HC}}$ is explained in Section~\ref{sec:results}. To construct $\hat H_\text{HC}$, we define a set of local transformations between spin configurations that act as discrete homotopies, connecting all and only those configurations that represent topologically equivalent | i.e., containing the same number of disjoint closed paths with the same nesting structure | ensembles of loops on the lattice. Leveraging the recently introduced framework of quantum equational reasoning~\cite{quantum_eq_reasoning}, from these transformations we build a local Hamiltonian whose ground space is an equal-amplitude superposition for each class of topologically equivalent configurations. We then introduce an additional local term that lifts the energy of all sectors corresponding to configurations with more than one loop, thereby restricting the ground-state subspace of the full parent Hamiltonian $\hat H_\text{HC}$ to the single, coherent quantum sample $\ket{X_\text{HC}}$.

The state $\ket{X_\text{HC}}$ encodes an infinite temperature ensemble in which all configurations are assigned equal probability. In Section~\ref{sec:finite_temp}, we show how imaginary time evolution~\cite{mcardle2019variational, Motta2020, PRXQuantum.3.010320,gluza2025doublebracketquantumalgorithmsquantum} can transform this state into a coherent quantum sample corresponding to an arbitrary temperature Boltzmann distribution of polymers. The approach is further extended to the thermodynamics of heteropolymers, where the spatial arrangement of distinct monomers determines the energy of each configuration. To this end, we design a quantum circuit that dresses the vertices of each Hamiltonian path with a target sequence of monomers, thereby mapping the state $\ket{X_\text{HC}}$ to a coherent quantum sample of heteropolymer configurations.

Finally, in Section~\ref{sec:tn}, we approximate the ground state of $\hat{H}_\text{HC}$ as a matrix product state~\cite{RevModPhys.77.259, Montangero2018}, thereby establishing a quantum-inspired framework to generate a compressed representation of the entire ensemble of closed Hamiltonian paths. The resulting state obeys an area law for entanglement entropy, which enables an efficient matrix-product-state approximation when one lattice dimension is fixed. Under these conditions, tensor network contractions can be used to efficiently compute expectation values, determine the partition function, and evaluate the probability of arbitrary configurations without resorting to sampling.

\section{Construction of the parent  Hamiltonian}\label{sec:results}

Here, we construct a local, frustration-free Hamiltonian  $\hat H_\text{HC}$  whose unique ground state $\ket{X_\text{HC}}$ is a coherent quantum sample encoding the complete set of closed Hamiltonian paths, named Hamiltonian cycles, on a rectangular lattice.

The construction relies on a set of local constraints that enforce the Hamiltonian condition, together with a family of local transformations that relate configurations that encode topologically equivalent paths. These ingredients naturally lead to a decomposition of the configuration space into topological sectors and allow us to analyze the spectral properties of $\hat H_\text{HC}$.

\subsection{Hamiltonian cycles as binary configurations}\label{subsec:configurations}

\begin{figure}
    \centering
    \includegraphics[width=\linewidth]{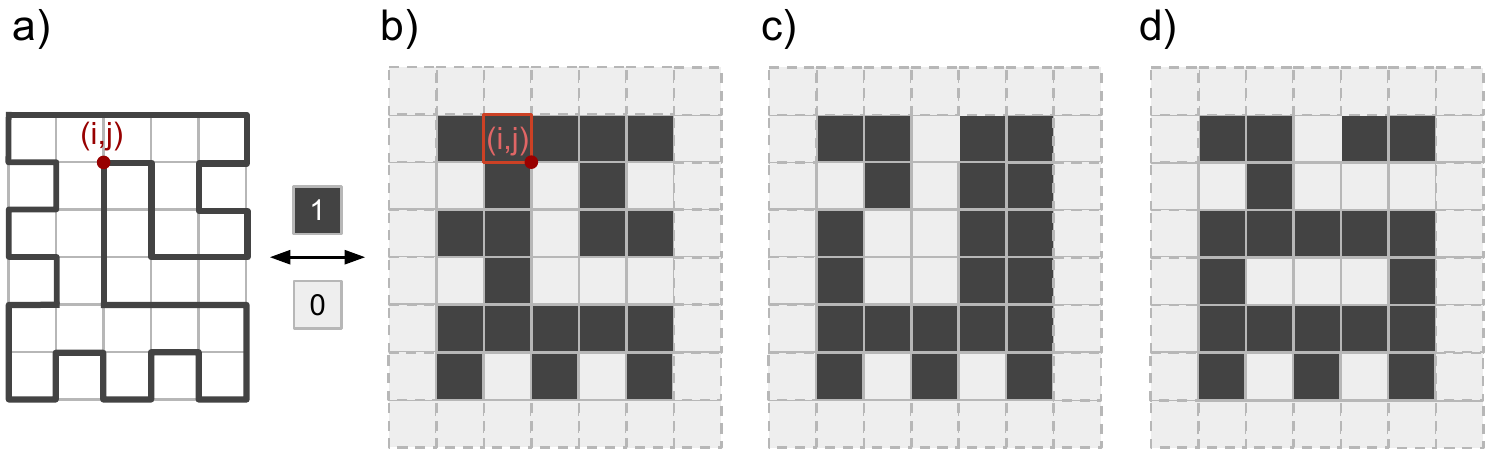}
    \caption{\textit{Hamiltonian cycles and multiloop configurations over lattices and dual lattices.} \textbf{Panel~(a):}~Hamiltonian cycle over a rectangular lattice.\textbf{ Panel~(b):}~configuration of the dual lattice encoding the Hamiltonian cycle in Panel~(a). \textbf{Panel~(c):}~configuration of the dual lattice encoding a multiloop violating the Hamiltonian condition: some vertex is never visited, and some vertex is visited two times. \textbf{Panel~(d):}~configuration of the dual lattice encoding a 2-factor violating the single-loop condition, as it consists of two disjoint loops.
    }
    \label{fig:different_curves}
\end{figure}

As a first step, we map Hamiltonian cycles to configurations of a binary system. We only consider graphs that can be embedded on a two-dimensional surface. By the Jordan curve theorem~\cite{jordan1893cours}, any closed path on such a graph separates the surface into two disjoint regions: an interior, namely the region enclosed by the path, and an exterior.

We now focus on the case of rectangular lattices with $n \times m$ vertices. A Hamiltonian cycle is a closed path that visits each vertex exactly once. To translate this notion into a binary representation, we introduce the \emph{dual lattice}, whose sites correspond to the $(n-1) \times (m-1)$ plaquettes of the square lattice — namely, the elementary regions into which the embedding subdivides the surface. In this picture, a Hamiltonian cycle is uniquely specified by the set of plaquettes lying in its interior: the boundary between interior and exterior plaquettes traces exactly the closed path on the original lattice. This construction gives rise to a natural binary encoding: for each dual plaquette we assign the value $1$ if it lies inside the cycle and $0$ otherwise (see Fig.~\ref{fig:different_curves}, Panels~(a) and~(b)).

For convenience, we extend the dual lattice by adding a fixed boundary frame of plaquettes in the state $0$. This modification is not essential to the construction but simplifies the definition of the operators introduced below. Under this convention, the dual lattice contains $(n+1) \times (m+1)$ plaquettes, and each vertex $(i,j)$ of the original lattice is surrounded by the four dual plaquettes $[[(i,j),(i,j+1)],[(i+1,j),(i+1,j+1)]]$ (see Fig.~\ref{fig:different_curves}, Panels~(a) and~(b)).

Each Hamiltonian cycle corresponds to a binary configuration of the dual lattice. However, the converse does not hold: a generic dual configuration represents a \emph{multiloop}, namely a collection of closed paths that may intersect or overlap on the original lattice (see Fig.~\ref{fig:different_curves}). To single out configurations corresponding to Hamiltonian cycles, two independent conditions must be met. The first is the \textit{Hamiltonian condition}, which enforces only those configurations that visit every vertex of the lattice exactly once (see Fig.~\ref{fig:different_curves}, Panel~c for a counterexample). Such configurations form collections of disjoint cycles, and are known as 2-factors in graph theory~\cite{harary2018graph} and as melts of rings in polymer physics~\cite{PhysRevLett.127.080501,doi:10.1126/sciadv.adi0204}. The second condition is the \textit{single-loop condition}, which excludes configurations composed of multiple disconnected cycles and retains only those forming a single closed path (see Fig.~\ref{fig:different_curves}, Panel~d for a counterexample). Unlike the Hamiltonian condition, the single-loop condition concerns the connectivity of the multiloop. Since graph connectivity is not definable in first-order logic, it cannot be verified using any collection of purely local Boolean constraints.~\cite{libkin2004elements}. Consequently, we cannot define a local classical Hamiltonian encoding all the Hamiltonian cycles in its degenerate ground space, but, as we show in here, we can encode a coherent superposition of all the Hamiltonian cycles in the unique ground state of a suitable quantum system.

We proceed as follows. Each classical configuration of the dual lattice
\begin{equation}
    \bm{\sigma}=(\sigma_{1,1},\dots,\sigma_{m+1,n+1})\in\Sigma=\{-1,1\}^{(m+1)(n+1)}
\end{equation}
is mapped to a computational basis state $\ket{\bm{\sigma}}$, and define the corresponding Hilbert space $\mathcal{H}$ spanned by these basis states. Then we define a local, non-diagonal Hamiltonian $\hat H_{\text{HC}}$ on $\mathcal{H}$ whose unique ground state is the equal-amplitude superposition of all Hamiltonian cycles:
\begin{equation}
    \ket{X_\text{HC}} :=\sum_{\bm{\sigma}\in X_\text{HC}} \frac{1}{\sqrt{|X_\text{HC}|}}\ket{\bm{\sigma}},
\end{equation}
where $X_{\text{HC}}$ is the set of all the dual-lattice configurations representing Hamiltonian cycles, and $|X_{\text{HC}}|$ is its cardinality. 

\begin{figure}
    \centering
    \includegraphics[width=\linewidth]{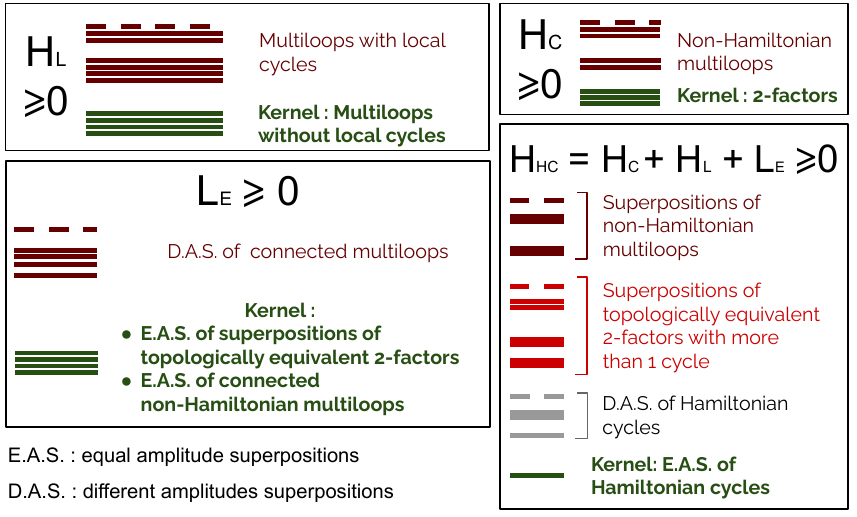}
    \caption{\textit{Schematic construction of the parent Hamiltonian $\hat H_\text{HC}$.} The Hamiltonian is expressed as the sum of three local positive semidefinite Hamiltonians, each enforcing a distinct condition on the states in its kernel. The ground state of $\hat H_\text{HC}$ simultaneously lies in these three kernels.} \label{fig:hamiltonian_construction}
\end{figure}

We construct $\hat H_{\text{HC}}$ as the sum of three local Hamiltonians:
\begin{equation}
    \hat H_{\text{HC}} = \hat H_\text{C} + \hat L_\text{E} + \hat H_\text{L}\;,
\end{equation}
where each term is frustration-free, and positive semi-definite, and enforces a distinct constraint on its ground states:
\begin{itemize}
    \item $\text{Ker}(\hat H_\text{C})$ spans states representing 2-factors;
    \item $\text{Ker}(\hat L_\text{E})$ spans states assigning equal amplitude to topologically equivalent 2-factors; and
    \item $\text{Ker}(\hat H_\text{L})$ spans states containing no cycles enclosing a single plaquette. We name these cycles local loops.
\end{itemize}
We will demonstrate that the kernel of $\hat H_{\text{HC}}$ is spanned only by $\ket{X_{\text{HC}}}$, which is the only state that simultaneously satisfies all constraints. It follows that the total Hamiltonian is frustration free.

The remainder of this section is devoted to defining $\hat H_\text{C}$, $\hat L_\text{E}$, and $\hat H_\text{L}$, and characterizing their spectra and eigenstates (Fig.~\ref{fig:hamiltonian_construction}).

\subsection{Definition of the Hamiltonian $\hat H_\text{C}$ enforcing $2$-factor configurations}\label{subsec:HC}

\begin{figure}
    \centering
    \includegraphics[width=\linewidth]{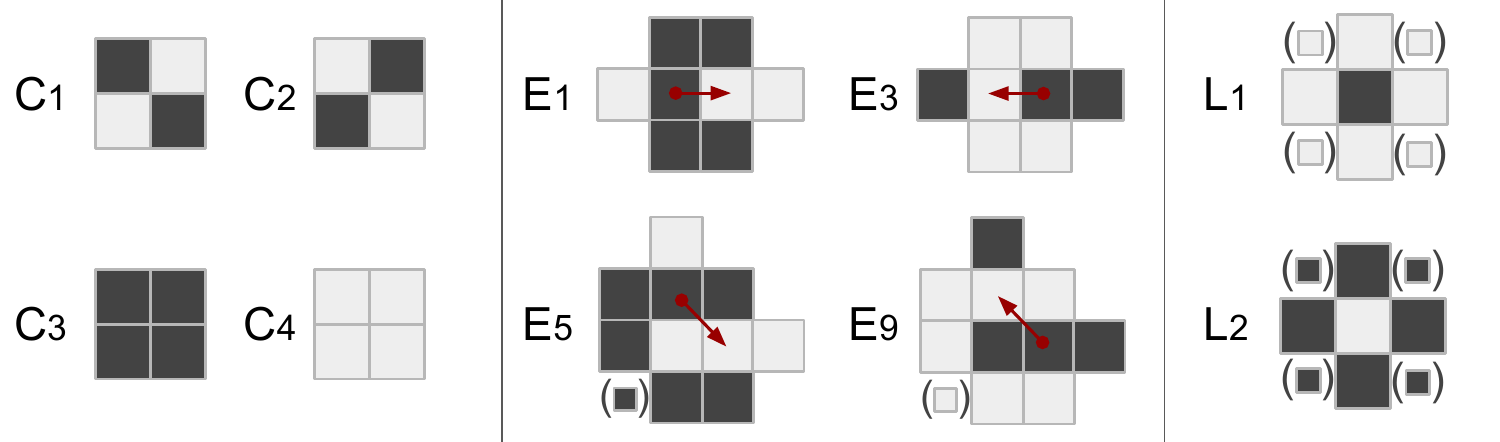}
    \caption{\textit{Local constraints and transformation rules involved in the definition of the Hamiltonian $\hat{H}_\text{HC}$.} \textbf{Panels~$C_1$, $C_2$, $C_3$, $C_4$:} local configurations forbidden in 2-factors. \textbf{Panels~$E_1$, $E_3$, $E_5$, $E_9$:} local transformations connecting the complete set of topologically equivalent 2-factors. The red arrow indicates the swap between two different color plaquettes, while the state of the other plaquettes remains unchanged. The missing transformations $\{E_2,E_4\}$ are obtained by rotating transformations $E_1$ and $E_3$ by 90 degrees. The missing transformations $\{E_6,E_7,E_8,E_{10},E_{11},E_{12}\}$ are obtained by rotating transformations $E_5$ and $E_9$ by 90, 180 and 270 degrees. Plaquettes shown in parentheses are not part of the corresponding rule or constraint, but are fixed in the displayed configuration as a consequence of the constraints $C_1$, $C_2$, $C_3$, and $C_4$. \textbf{Panels~$L_1$, $L_2$:} local configurations encoding local cycles.
    }
    \label{fig:constraints_and_rules}
\end{figure}

Violations of the Hamiltonian condition can be ruled out by enforcing purely local constraints. A vertex $(i,j)$ of the original lattice is traversed exactly once if and only if the four dual plaquettes surrounding it, $(i,j),(i,j+1),(i+1,j),(i+1,j+1)$, avoid the forbidden configurations $C_1$, $C_2$, $C_3$, and $C_4$ shown in Fig.~\ref{fig:constraints_and_rules}. Each of these patterns corresponds to an invalid local arrangement in which the degree of vertex $(i,j)$, that is, the number of path edges incident on the vertex, differs from two. Hence, identifying 2-factors on the original lattice reduces to finding binary configurations of the dual lattice that simultaneously satisfy a set of local four-plaquette constraints excluding the forbidden configurations $C_1$, $C_2$, $C_3$, and $C_4$.

We map each forbidden four-plaquette pattern to a corresponding local diagonal operator, whose expecation value returns $1$ to configurations violating the constraint within the block of plaquettes, and $0$ to all others:
\begin{align}
    C_1\rightarrow\hat C^{[1]}_{ij} =& \widehat{P}^{[0]}_{ij}\widehat{P}^{[0]}_{i,j+1}\widehat{P}^{[0]}_{i+1,j}\widehat{P}^{[0]}_{i+1,j+1}\nonumber\\
    C_2\rightarrow\hat C^{[2]}_{ij} =&\widehat{P}^{[1]}_{ij}\widehat{P}^{[1]}_{i,j+1}\widehat{P}^{[1]}_{i+1,j}\widehat{P}^{[1]}_{i+1,j+1}\nonumber\\
    C_3\rightarrow\hat C^{[3]}_{ij} =& \widehat{P}^{[0]}_{ij}\widehat{P}^{[1]}_{i,j+1}\widehat{P}^{[1]}_{i+1,j}\widehat{P}^{[0]}_{i+1,j+1}\nonumber\\
    C_4\rightarrow\hat C^{[4]}_{ij} =& \widehat{P}^{[1]}_{ij}\widehat{P}^{[0]}_{i,j+1}\widehat{P}^{[0]}_{i+1,j}\widehat{P}^{[1]}_{i+1,j+1}\;,
\end{align}
where $\widehat{P}^{[1]}_{ij} := \left(\ket{1}\bra{1}\right)_{ij}$ and  $\widehat{P}^{[0]}_{ij} := \left(\ket{0}\bra{0}\right)_{ij}$. The sum of these operators yields the Hamiltonian
\begin{align}\label{eq:hc}
    \hat H_\text{C} = \sum_{\substack{1\leq i \leq m\\ 1\leq j\leq n}} \left(C^{[1]}_{ij}+C^{[2]}_{ij}+C^{[3]}_{ij}+C^{[4]}_{ij}\right)\;,
\end{align}
which is diagonal in the computational basis and counts the number of violations of the constraints. By construction, the ground states are degenerate, have zero energy, and encode all possible 2-factors.

\subsection{Topological deformation rules}\label{subsec:deformation_rules}

The single-loop condition is a topological property that depends on the 
 global structure of the loops
and cannot be detected by any finite set of local, diagonal constraints.
Nevertheless, configurations that are \emph{topologically equivalent} | i.e., configurations containing the same number of loops with the same nesting structure | can be transformed into one another through sequences of local deformations that neither cut nor merge loops. These deformations act as discrete analogues of homotopies, then preserving both the number of cycles and their nesting structure.

Building on this observation, we define a family of transformation rules that locally modify dual-lattice configurations. The full set of rules is:
\begin{equation}
E = \bigcup_{\substack{k\in[1,12]\\1\leq i\leq m+2-w_k\\1\leq j\leq n+2-h_k}} \left\{e_{ij}^{[k]}, {e_{ij}^{[k]}}^{-1}\right\}\;,
\end{equation}
where
\begin{equation}
e^{[k]}_{ij} : \bm \sigma \in \Sigma \rightarrow \bm \sigma' \in \Sigma\;,
\end{equation}
is a transformation rule acting on a block of $w_k\times h_k$ plaquettes whose upper–left corner is $(i,j)$, and ${e_{ij}^{[k]}}^{-1}$ is its inverse. Each rule is defined as follows: if the block matches the pattern shown in Panels~$E_k$ of Fig.~\ref{fig:constraints_and_rules}, the rule swaps a black and a white plaquette along the red arrow, leaving all other plaquettes unchanged; otherwise, the rule is undefined on $\bm\sigma$.

The set $E$ satisfies two properties:
\begin{enumerate}[a)]
    \item \textit{soundness}: every rule maps any 2-factor into a topologically equivalent 2-factor; and
    \item \textit{completeness}: any pair of topologically equivalent 2-factors can be connected by a finite sequence of rules in $E$.
\end{enumerate}
In this sense, $E$ forms a \textit{sound and complete equational theory} for topologically equivalent 2-factors. Soundness follows because no rule generates any of the forbidden patterns $C_1$, $C_2$, $C_3$, and $C_4$ thereby preserving the Hamiltonian condition; moreover, each rule moves a plaquette to a region of the same color, deforming the loops without cutting or merging them, therefore preserving topology. Previous works~\cite{ham_reconf_rect, ham_reconf_l_shape, nishat2020reconfiguration} demonstrated the completeness of $E$ for any pair of Hamiltonian cycles. Here, we conjecture that completeness also extends to any pair of topologically equivalent 2-factors.

It is important to note that the equivalence relation generated by 
$E$ coincides with topological equivalence only within the subspace of 2-factors. For example, since all rules in $E$ conserve the number of plaquettes in the state  $\sigma_{ij}=1$, they cannot connect two cycles enclosing regions of different areas, even though such cycles are topologically equivalent.

The configuration space $\Sigma$ is partitioned into equivalence classes containing all and only elements that are connected by the rules in $E$. We name this partition $S$:
\begin{align}\label{eq:conf_partition}
    &S =  \{ X_\text{HC}, X_1, \dots , X_l , Z_1 , \dots,  Z_m\},\nonumber\\
    &\Sigma = X_\text{HC}\cup X_1\cup\dots \cup X_l \cup Z_1\cup \dots\cup Z_m\nonumber\\
    &A\cap B=\varnothing \quad\forall\; A,B \in S\;,
\end{align}
where $X_\text{HC}$ contains all dual-lattice configurations encoding Hamiltonian cycles; each $X_\alpha$ contains configurations encoding topologically equivalent 2-factors with the same number of cycles $l_\alpha>1$ and the same nesting structure; and each $Z_\beta$ contains configurations violating the Hamiltonian condition but mutually connected by the rules in 
 $E$.

\subsection{From rules to operators}\label{subsec:rules_to_ops}

Each rule $e \in E$ is naturally associated with an operator $\hat e$ encoding its action on the computational basis of 
$\mathcal{H}$. For a rule $e^{[k]}_{ij}$, defined on a block of plaquettes with upper-left corner $(i,j)$, the corresponding operator is
\begin{equation}
    \hat e^{[k]}_{ij} = \sum_{\bm{\sigma}}\ket{e^{[k]}_{ij}(\bm{\sigma})}\bra{\bm{\sigma}}\;,
\end{equation}
and the operator encoding the inverse rule is
\begin{equation}
    {\hat e^{[k]\dag}_{ij}} = \sum_{\bm{\sigma}}\ket{\bm{\sigma}}\bra{e^{[k]}_{ij}(\bm{\sigma})} = \sum_{\bm\sigma}\ket{e_{ij}^{[k]-1}(\bm{\sigma})}\bra{\bm{\sigma}}\;,
\end{equation}
where the sums run over all configurations $\bm\sigma$ for which the rule is defined, i.e., those matching the pattern in Panel~$E_k$ of Fig.~\ref{fig:constraints_and_rules}. We denote the set of rule operators as
\begin{equation}
    \mathcal{E} = \{\hat e \;|\;e\in E\}\;.
\end{equation}
The explicit form of the operators $\hat e$ is shown in Appendix~\ref{app:rules}. We note that the operators inherit locality from the corresponding rules.

Since each configuration in $\Sigma$ corresponds to a computational basis state,  following Eq.~\ref{eq:conf_partition}, we can decompose the Hilbert space as a direct sum of corresponding orthogonal subspaces:
\begin{equation}\label{eq:equivalence_partition}
    \mathcal{H} = \mathcal{X}_\text{HC}\oplus \mathcal{X}_1\oplus\dots \oplus \mathcal{X}_l \oplus \mathcal{Z}_1\oplus \dots\oplus \mathcal{Z}_m
\end{equation}
where we define$\mathcal{A} = \text{span}\left(\{\ket{\sigma}\;|\;\sigma\in A\}\right)$ for any set $A \in S$. The action of the rules in $E$ generates each set $A\in S$, thus $\mathcal{A}$ is the linear span of all states reachable from any $\ket{\sigma} \in A$ by applying the rule operators:
\begin{equation}\label{eq:orbits_span}
    \mathcal{A} = \mathrm{span} \Big\{ 
\hat e_{i_1} \cdots \hat e_{i_m} \,|\bm\sigma\rangle 
\;\Big|\; m \in \mathbb{N},\; \hat e_{i_j} \in \{\mathcal{E}\}\Big\}.
\end{equation}

\subsection{Definition of a Hamiltonian $\hat L_\text{E}$ enforcing equivalence between $2$-factors.}\label{subsec_le}

Following the quantum equational reasoning framework in Ref.~\cite{quantum_eq_reasoning}, we define the Laplacian Hamiltonian
\begin{equation}
\hat L_\text{E} = \sum_{k=1}^{12}\;\sum_{\substack{1\leq i\leq m+2-w_k\\ 1\leq j\leq n+2-h_k}} \left[\left(\hat{e}^{[k]}_{ij} + h.c.\right)^2 - \left(\hat{e}^{[k]}_{ij} + h.c.\right)\right]\;,
\end{equation}
where $\hat{e}^{[k]}_{ij}$ denotes the local operator implementing rule $E_k$, and $w_k \times h_k$ is the support of that rule.

The Hamiltonian $\hat L_\text{E}$ is local and frustration-free. In the computational basis, $\hat L_\text{E}$ is the discrete Laplacian of the graph $G$ whose vertices represent dual-lattice configurations and whose edges correspond to applications of the rules in $E$. The graph $G$ is disconnected, and each connected subgraph is an equivalence class $A \in S$. Following the decomposition of $\mathcal{H}$, the block-diagonal form of the Laplacian Hamiltonian is:
\begin{equation}\label{eq:le_decomposition}
    \hat L_\text{E} = \hat L_\text{E}|_{\mathcal{X}_\text{HC}}\oplus \dots\oplus \hat L_\text{E}|_{\mathcal{X}_{l}}\oplus \dots\oplus \hat L_\text{E}|_{\mathcal{Z}_{m}}\;,
\end{equation}
where $\hat L_\text{E}|_{\mathcal{A}}$ is the Laplacian of the connected subgraph $G_A$.
The ground state of each block is the zero-energy state given by the equal-amplitude superposition of all computational-basis states in the corresponding class:
\begin{equation}
    \ket{A} := \textit{G.S.}\left(\hat L_\text{E}|_{\mathcal{A}}\right) = \sum_{\bm\sigma\in A}\frac{1}{\sqrt{|A|}}\ket{\bm\sigma}\;.
\end{equation}
The target state $\ket{X_\text{HC}}$ is the ground state that lies in $\mathcal{X}_\text{HC}$.

Within each subspace, the excited states are superpositions whose amplitudes satisfy discrete Laplace equations on the corresponding connected subgraph $G_A$. Thus, they are a graph generalisation of harmonic functions. The first excited state defines a bipartite clustering of $G_A$~\cite{spectral_clustering_0,vonLuxburg2007} and the corresponding energy gap $\Delta_\mathcal{A}$ may signal the presence of bottlenecks in $G_A$~\cite{spectral_clustering_0}. Finally,  the mixing time $\tau_A$ of a random walk on $G_A$ is inversely proportional to $\Delta_\mathcal{A}$~\cite{montenegro}.

\subsection{Definition of a Hamiltonian $\hat H_\text{L}$ enforcing no local cycles.}\label{subsection:hl}

Here, we construct the Hamiltonian $\hat H_\text{L}$, whose kernel spans the configurations that do not contain any \emph{local cycle}. Local cycles correspond to the patterns $L_1$ and $L_2$ in Fig.~\ref{fig:constraints_and_rules}, each representing a loop enclosing a single plaquette. 

As in the construction of $\hat H_\text{C}$, we associate to each forbidden pattern a local diagonal operator whose expectation value returns $1$ over configurations with a local cycle, and $0$ otherwise. For the block of plaquettes with upper-left corner $(i,j)$, we define:
\begin{align}
    L_1&\rightarrow \hat L_{ij}^{[1]} = \widehat{P}^{[0]}_{i,j+1}\widehat{P}^{[0]}_{i+1,j}\widehat{P}^{[1]}_{i+1,j+1}\widehat{P}^{[0]}_{i+1,j+2}\widehat{P}^{[0]}_{i+2,j+1}\nonumber\\
    L_2&\rightarrow \hat L_{ij}^{[2]} = \widehat{P}^{[1]}_{i,j+1}\widehat{P}^{[1]}_{i+1,j}\widehat{P}^{[0]}_{i+1,j+1}\widehat{P}^{[1]}_{i+1,j+2}\widehat{P}^{[1]}_{i+2,j+1}.
\end{align}
Summing over all positions gives the local Hamiltonian:
\begin{equation}\label{eq:hl}
\hat H_\text{L} = \sum_{\substack{1\leq i \leq m-1\\ 1\leq j\leq n-1}} \Big[ \hat L_{ij}^{[1]} + \hat L_{ij}^{[2]}\Big]\;.
\end{equation}
Ground states of $\hat H_\text{L}$ have zero energy and correspond to configurations without local cycles, i.e.~no plaquette is enclosed by a unit-area loop. Configurations containing local cycles appear as excited states, with energy penalty proportional to the number of such cycles, and a minimum gap $\Delta_{\text{L}}=1$. As we show in the next section, disjoint cycles enclosing more than a single plaquette are also penalized through the interplay between $\hat H_\text{L}$ and $\hat L_\text{E}$.

\subsection{Spectral properties of $\hat H_\text{HC}$}\label{subsec:spectral}

\begin{figure}
    \centering
    \includegraphics[width=\linewidth]{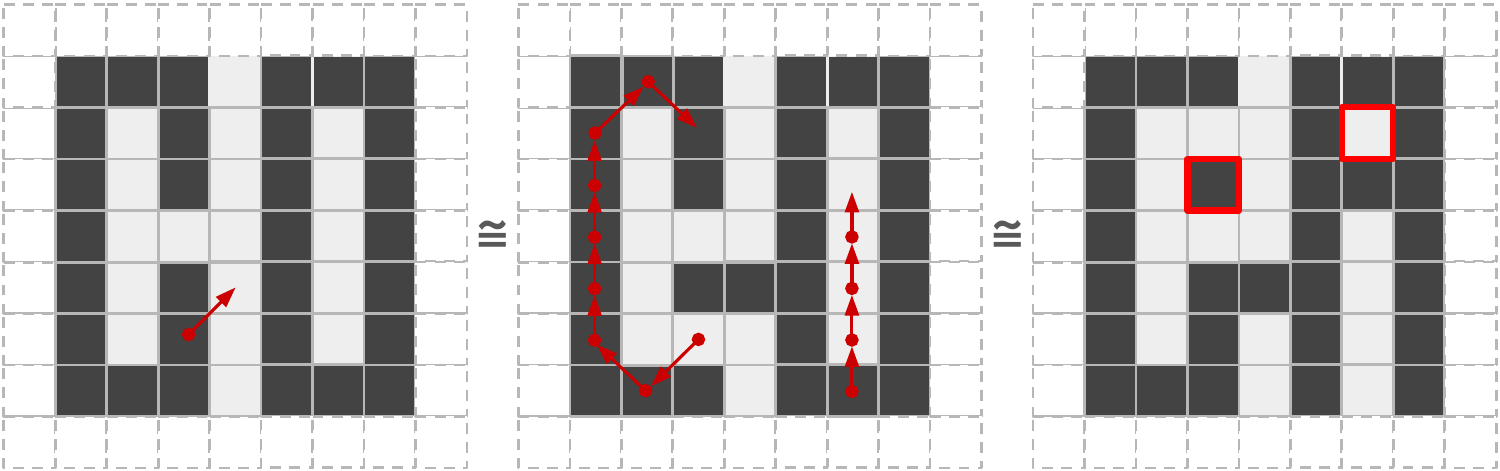}
    \caption{\textit{Contraction of a loop to a local loop.} Rules in $E$ can be used to contract a 2-factor with more than one cycle to a 2-factor where innermost cycles enclose a single plaquette of the dual lattice.}
    \label{fig:contract_to_local_loops}
\end{figure}

We now analyse the spectral properties of the total Hamiltonian defined as:
\begin{equation}
\hat H_\text{HC} = \hat H_\text{C} + \hat H_\text{L} + \hat L_\text{E}\;.
\end{equation}
\paragraph{Block decomposition of $\hat H_\text{C}$.} The Hamiltonian $\hat H_\text{C}$ is block-diagonal in the space decomposition in Eq.~\ref{eq:equivalence_partition}. Because the rules in $E$ preserve the Hamiltonian condition, i.e.~$[\hat C^{[k]}_{ij},\hat e^{[l]}_{i'j'}] = 0$, $\hat H_{C}$ is constant on each subspace:
\begin{align}\label{eq:hc_decomposition}
    \hat H_{C} =& \hat H_\text{C}|_{\mathcal{X}_\text{HC}}   \oplus\dots\oplus \hat H_\text{C}|_{\mathcal{X}_{l}}\oplus \dots\oplus \hat H_\text{C}|_{\mathcal{Z}_{m}}\nonumber\\
     = &0\cdot\id \oplus \dots  \oplus 0\cdot\id  \oplus \dots \oplus N_{Z_m}\cdot \id\;,
\end{align}
where no violation of the Hamiltonian condition happens in the equivalence classes $\mathcal{X}_\text{HC},\dots,\mathcal{X}_l$ encoding 2-factors, and $N_{Z_i}\geq 1$ counts the violations in any configuration in $\mathcal{Z}_{i}$. 

\paragraph{Block decomposition of $\hat H_\text{L}$.} Being itself diagonal, $\hat H_\text{L}$ can be written with the same block structure:
\begin{equation}\label{eq:hl_decomposition}
    \hat H_\text{L} =\hat H_\text{L}|_{\mathcal{X}_\text{HC}}\oplus \dots\oplus \hat H_\text{L}|_{\mathcal{X}_{l}}\oplus \dots \oplus \hat H_\text{L}|_{\mathcal{Z}_{m}}\;.
\end{equation}
Since Hamiltonian cycles contain no local cycles, $\hat H_\text{L}|_{\mathcal{X}_\text{HC}}= 0\cdot\id$. In contrast, by applying the deformation rules in $E$, any multi-loop 2-factor can be transformed into a configuration where at least one innermost cycle encloses a single plaquette, see Fig.~\ref{fig:contract_to_local_loops}. Thus, each equivalence class $\mathcal{X}_{\alpha}$ with $l_{\alpha}>1$ | i.e., containing a 2-factor with more than one loop | includes configurations with a local cycle. As a consequence, $\hat H_\text{L}$ is positive in each $\mathcal{X}_l$ with $l_\alpha >1 $.

Combining Eqs.~\ref{eq:le_decomposition},~\ref{eq:hc_decomposition}, and~\ref{eq:hl_decomposition}, we can block-diagonalize the total Hamiltonian:
\begin{align}\label{eq:hhc_decomposition}
    \hat H_\text{HC} &= \hat L_\text{E}|_{\mathcal{X}_\text{HC}}\oplus\dots\oplus \left(\hat L_\text{E} + \hat H_\text{L}\right)\Big|_{\mathcal{X}_{l}}\nonumber\\
    &\oplus \dots \oplus \left(N_{Z_m}\cdot\id+\hat L_\text{E} + \hat H_\text{L}\right)\Big|_{\mathcal{Z}_{m}}\;.
\end{align}

From the last decomposition, the spectral structure is thus clear:
\begin{itemize}
    \item in the subspace $\mathcal{X}_\text{HC}$, the ground state is $\ket{X_\text{HC}}$, the first energy gap is the Laplacian gap $\Delta_{\mathcal{X}_\text{HC}}$, and excited states are harmonic superpositions of Hamiltonian cycles.
    \item Subspaces $\mathcal{X}_\alpha$ span superpositions of topologically equivalent 2-factors with more than one cycle, which include configurations penalized by $\hat H_\text{L}$. The ground state of $\hat L_\text{E}|_{\mathcal{X}_\alpha}$, which is the equal-amplitude superposition $\ket{X_\alpha}$, acquires energy proportional to the fraction of equivalent configurations containing at least one local cycle. As we show in Appendix~\ref{app:energy_gap}, this fraction scales as $1/(n\cdot m)$, and, as a consequence of the Halmos’ two-subspaces theorem~\cite{halmos1969two,BOTTCHER20101412}, the ground state energy in the sector $\mathcal{X}_\alpha$ is lower bounded by $\frac{\Delta_{\mathcal{X}_1}}{4\,m\; n\;(1+\Delta_{\mathcal{X}_1})}$.
    \item Finally, non-Hamiltonian cycles are encoded in the subspaces $\mathcal{Z}_i$. In these subspaces, the ground states have energy at least $N_{Z_i}\geq1$.
\end{itemize}

We conclude that $\hat H_\text{HC}$ is a local, frustration-free Hamiltonian with a unique zero-energy ground state $\ket{X_{\text{HC}}}$. The first excited energy in each block is controlled either by the Laplacian gap of that block and, in the case of multiloop sectors $\mathcal{X}_\alpha$, by the probability of encountering a local cycle within the class. The overall lower bound on the first energy gap reads:
\begin{equation}\label{eq:gap}
\Delta\geq\min\left(\Delta_{\mathcal{X}_\text{HC}},\frac{\Delta_{\mathcal{X}_1}}{4mn(1\!+\!\Delta_{\mathcal{X}_1})},\dots,\frac{\Delta_{\mathcal{X}_l}}{4mn(1\!+\!\Delta_{\mathcal{X}_l})},1\right)\;.
\end{equation}

\section{Polymers and heteropolymers at finite temperature}\label{sec:finite_temp}

\begin{figure*}[t]
    \centering
    \includegraphics[width=\textwidth]{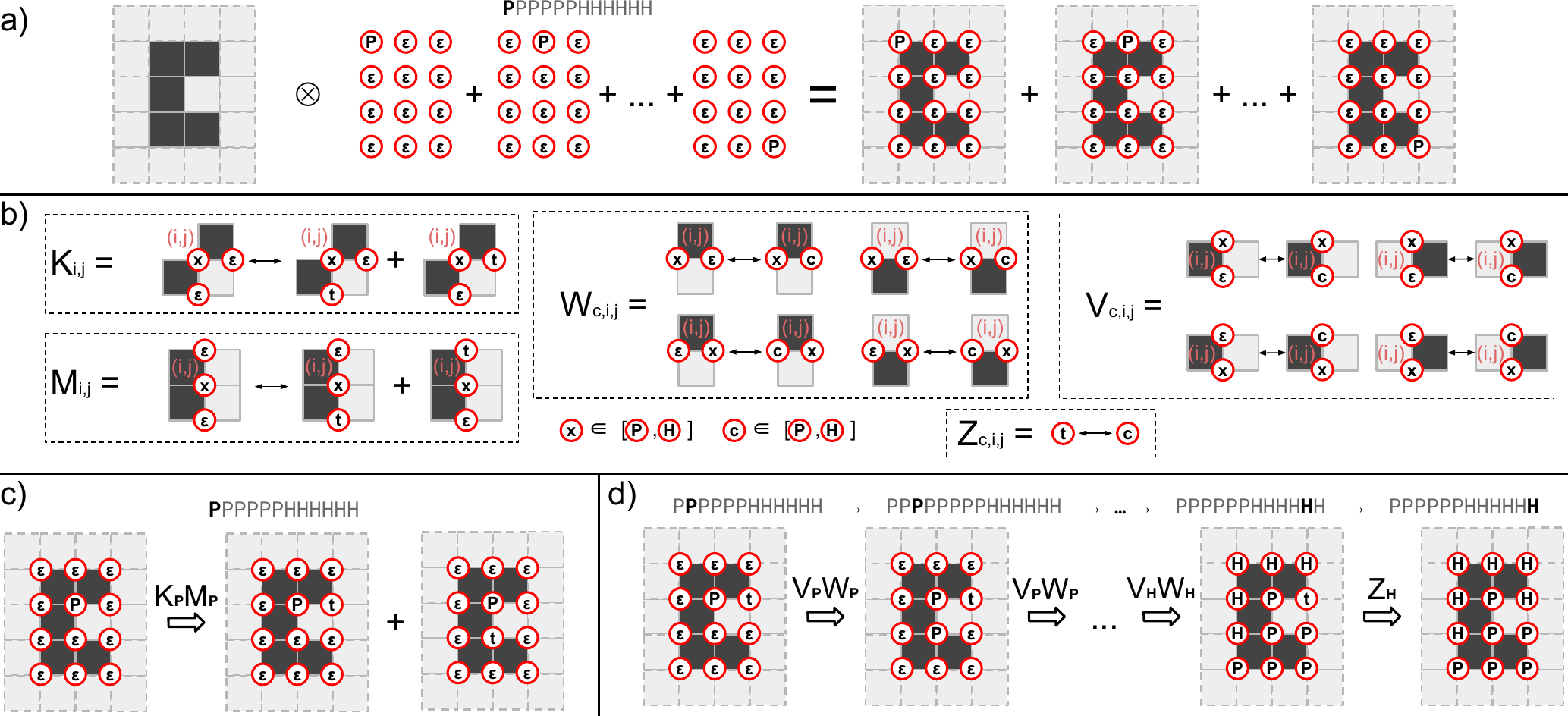}
    \caption{ \textit{Dressing Hamiltonian cycles with a monomer sequence.} Monomers are represented by the characters \textit{P} and \textit{H}, and the sequence is \textit{PPPPPPHHHHHH}. \textbf{Panel~(a)}: A qudit is assigned to each lattice vertex, with its internal state encoding either a monomer character, the empty character, or the terminal character. The qudits are initially prepared in a $W$ state, representing a superposition of all possible placements of the first monomer type of the sequence.
    \textbf{Panel~(b)}: Action of the operators $\hat K_{i,j}$, $\hat M_{i,j}$, $\hat W_{c,i,j}$, $\hat V_{c,i,j}$, and $\hat Z_{c,i,j}$ on local configurations. The action of $\hat K_{i,j}$ and $\hat M_{i,j}$ is identical for rotated configurations and for configurations in which plaquettes in state $0$ are exchanged with plaquettes in state $1$, and vice versa. For all other local configurations, the operators act as the identity. \textbf{Panel~(c)}: Action of $\prod_{i,j}\hat K_{i,j}\hat M_{i,j}$ on a configuration containing a single monomer. 
    \textbf{Panel~(d)}: Repeated application of the sequence-extension operators $\prod_{ij}\hat K_{ij}\hat M_{i,j}$ and the final action of $\prod_{ij}\hat Z_{c_{mn},i,j}$, completing the dressing of the Hamiltonian cycle with the full monomer sequence.}
    \label{fig:glob_protein}
\end{figure*}

If the energy $E_\text{poly}(\bm\sigma)$ of a maximally compact polymer depends only on its spatial configuration $\bm \sigma$, then the state $\ket{X_\text{HC}}$ represents the corresponding infinite temperature Boltzmann distribution. After preparing $\ket{X_\text{HC}}$, amplitude amplification enables accelerated estimation of classical expectation values over the ensemble of compact polymers, but also the estimation of quantities expressible as expectation values of non-diagonal operators, such as the partition function (see Appendix~\ref{app:counting}). Appendix~\ref{app:amplification} illustrates with an example the use of amplitude amplification for expectation value estimation, and Ref.~\cite{doi:10.1098/rspa.2015.0301} provides a comprehensive overview of this methodology.

To generalize our approach to finite temperature, it is necessary to prepare a coherent quantum sample of the Boltzmann distribution $p(\bm\sigma)=\sum_{\bm\sigma \in X_\text{HC}}e^{-\beta E_\text{poly}(\bm\sigma)}/Z(\beta)$, where $Z(\beta)=\sum_{\bm\sigma \in X_\text{HC}}e^{-\beta E(\bm\sigma)}$ is the partition function. The corresponding coherent quantum sample is
\begin{equation}
    \ket{Z(\beta)}:=\frac{1}{Z(\beta)}\sum_{\bm\sigma \in X_\text{HC}}\sqrt{e^{-\beta E(\bm\sigma)}}\ket{\bm\sigma}\;.
\end{equation}

This coherent sample can be prepared by evolving the state $\ket{X_\text{HC}}$ for an imaginary time $\beta/2$ using the classical Hamiltonian $\hat H_\text{poly}=\sum_{\bm\sigma}E_\text{poly}(\sigma)\ket{\bm\sigma}\bra{\bm\sigma}$. Implementing imaginary time evolution~\cite{mcardle2019variational, Motta2020, PRXQuantum.3.010320,gluza2025doublebracketquantumalgorithmsquantum} yields
\begin{equation}
    \frac{e^{-\frac{\beta}{2} \hat H_\text{poly}}\ket{X_\text{HC}}}{\left|\left|e^{-\frac{\beta}{2} \hat H_\text{poly}}\ket{X_\text{HC}}\right|\right|^2} = \frac{1}{Z(\beta)}\sum_{\bm\sigma \in X_\text{HC}}e^{-\frac{\beta}{2} E(\bm\sigma)}\ket{\bm\sigma}=\ket{Z(\beta)}\;.
\end{equation}

The Hamiltonian $\hat H_\text{poly}$ assigns an energy to each polymer configuration based solely on its shape, as represented by a Hamiltonian cycle. However, this model does not apply to heteropolymers, which are polymers composed of a sequence of chemically distinct monomer species. In heteropolymers, the configuration energy for a given sequence depends on both the spatial arrangement of monomers within a Hamiltonian cycle and on how the sequence is assigned to the latter. Therefore, generating a coherent quantum sample of heteropolymer configurations requires implementing imaginary-time evolution on an initial infinite temperature sample that encodes both the above features.

Thus, to encode compact heteropolymers, we need to dress the vertices of each Hamiltonian cycle with a sequence of symbols encoding monomers, starting from a designated reference edge. These symbols are called the chemical alphabet. In the lattice formulation considered here, we assign to each lattice vertex a qudit whose energy levels are labeled by the symbols of the chemical alphabet (see Fig.~\ref{fig:glob_protein}). A classical configuration of the full system is specified by a plaquette configuration $\ket{\bm{\sigma}}$, describing a Hamiltonian cycle, together with one of the corresponding qudit configurations $\ket{\bm{v}_{\bm{\sigma},i}}$. In each configuration $\ket{\bm{v}_{\bm{\sigma},i}}$, with $i\in[1,\dots,n\cdot m]$, the monomers are sequentially placed along the vertices of the cycle starting from one of the $n \times m$ vertices.

The remainder of this section is devoted to the construction of a quantum algorithm that dresses each of the configurations encoded in the state $\ket{X_\textbf{HC}}$ with the sequence of monomers specified by a given heteropolymer. This dressing process generates a coherent superposition $\ket{D}$ of maximally compact heteropolymer configurations, with information about both the Hamiltonian cycle and monomer type assignment:
\begin{align}
    \ket{D} = \frac{1}{\sqrt{n\cdot m|X_\text{HC}|}}\sum_{\bm{\sigma}\in X_\text{HC}} \sum_{i=1}^{n\cdot m}\ket{\bm{\sigma}}\otimes\ket{v_{\bm{\sigma},i}}\;.
\end{align}

As a first step, we extend the chemical alphabet | and consequently the number of internal qudit levels | by introducing two additional symbols, $\epsilon$ and $t$, which we denote as the empty character and the terminal character, respectively. Then, we prepare the subsystem encoding all the plaquettes in the state $\ket{X_\text{HC}}$, and the subsystem encoding all the vertices in a $W$ state
\begin{equation}
    \ket{W_{c_1}} = \frac{1}{\sqrt{m\cdot n}}\left(\ket{c_1\epsilon\dots\epsilon} + \ket{\epsilon c_1\epsilon\dots\epsilon} + \dots + \ket{\epsilon\dots\epsilon c_1}\right),
\end{equation}
where $c_1$ is the first character of the monomer sequence. The state of the resulting many-body system is
\begin{equation}
    \ket{X_\text{HC}}\otimes\ket{W_{c_1}} = \frac{1}{\sqrt{m\cdot n|X_\text{HC}|}}\sum_{\bm{\sigma}\in X_\text{HC}} \ket{\bm{\sigma}}\otimes\ket{W_{c_1}},
\end{equation}
where the elements in the sum are pictorially described in Figure~\ref{fig:glob_protein} Panel~(a).

We construct a circuit of local operators $\ket{X_\text{HC}}\otimes\ket{W_{c_1}}$ evolving the initial state to the state $\ket{D}$. To this aim, we introduce local many-body unitaries $\hat K_{i,j}$, $\hat M_{i,j}$, $\hat W_{c,i,j}$, $\hat V_{c,i,j}$, and $\hat Z_{c,i,j}$, whose action on the computational basis states of the local subsystem is depicted in Figure~\ref{fig:glob_protein} Panel~(b).

The operators $\hat K_{i,j}$ and $\hat M_{i,j}$ act on pairs of empty characters that are adjacent to a monomer character along the Hamiltonian cycle. They replace the state $\ket{\epsilon\epsilon}$ of the two empty vertices with the superposition $\left(\ket{t\epsilon}+\ket{\epsilon t}\right)/\sqrt{2}$. The action of these operators over the entire lattice is illustrated in Fig.~\ref{fig:glob_protein} Panel~(c). The resulting state is a quantum superposition of Hamiltonian configurations in which the first monomer of the sequence is placed at every possible vertex, while a terminal character appears either before or after the monomer, thereby encoding the two possible orientations of the sequence.

The next step of the procedure consists of sequentially adding all monomers of the sequence to each configuration. This is achieved by applying the operators $\hat W_{c,i,j}$ and $\hat V_{c,i,j}$, which place the monomer character $c$ on an empty vertex adjacent to an existing monomer along the Hamiltonian cycle. Acting with these operators over the entire lattice appends the character $c$ to the next empty site along the Hamiltonian cycle, as illustrated in Fig.~\ref{fig:glob_protein} Panel~(d).

Once all monomers except the final one have been placed, the operator $\hat Z_{c,i,j}$ is applied to each lattice vertex, replacing the terminal character $t$ with the last monomer of the sequence (see Fig.~\ref{fig:glob_protein} Panel~(d)).

In this way, we finally obtain a circuit capable of preparing the target state: 
\begin{align}
    \ket{D} =&  \prod_{ij}\hat Z_{c_{mn},i,j}\left[\prod_{l\in[m n-1,\dots,2]}\left(\prod_{ij}\hat W_{c_l,i,j}\hat V_{c_l,i,j}\right)\right]\nonumber\\
    &\prod_{ij}\hat K_{ij}\hat M_{i,j}\left(\ket{X_\text{HC}}\otimes\ket{W_{c_1}}\right).
\end{align}

\section{Tensor network applications}\label{sec:tn}

\begin{figure}
    \centering
    \includegraphics[width=\linewidth]{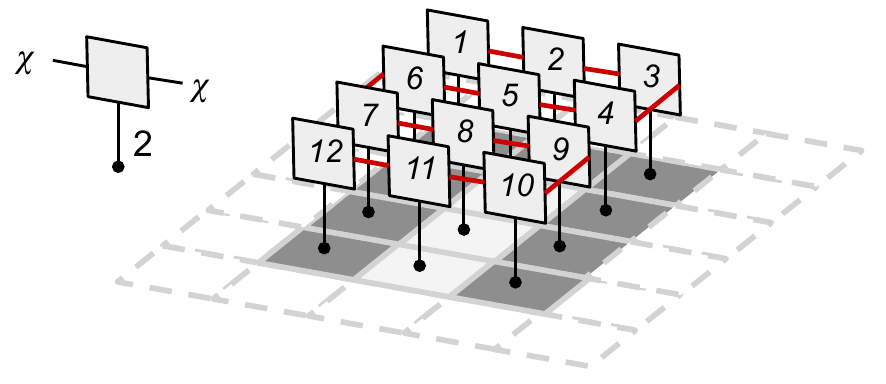}
    \caption{\textit{Encoding dual lattice configuration in a matrix product state.} Each plaquette in the bulk is associated with a rank-3 tensor. The tensors encoding plaquettes that are adjacent along a snake in the bulk are connected (red links).}
    \label{fig:zigzag}
\end{figure}

\begin{figure*}[t]
    \centering
    \includegraphics[width=\textwidth]{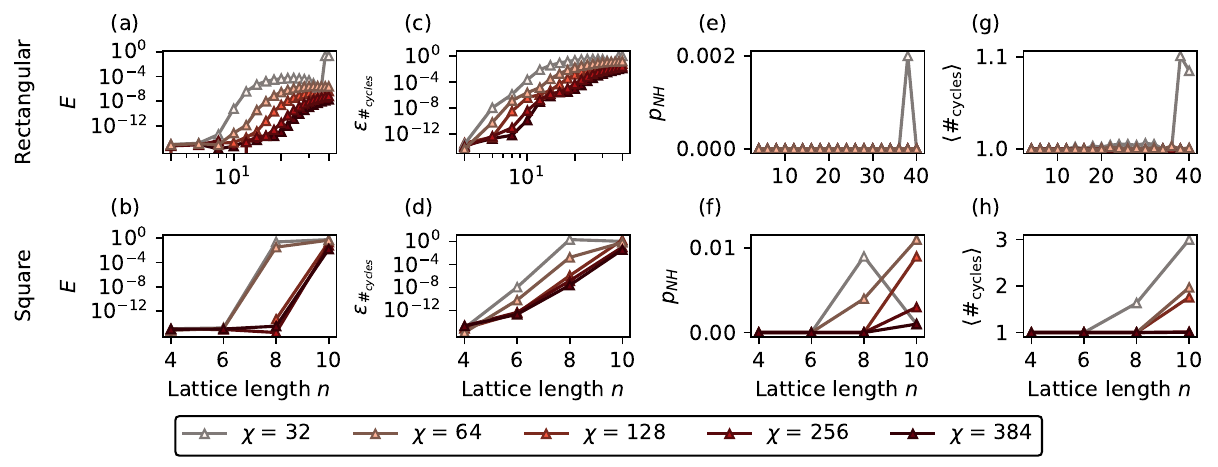}
    \caption{\textit{Quality of the MPS approximations of $\ket{X_\text{HC}}$ for different lattice lengths $n$ and bond dimensions $\chi$.} Panels~(a,c,e,g) correspond to rectangular lattices of size $6\times n$, while panels~(b,d,f,h) correspond to square lattices of size $n\times n$. \textbf{Panels~(a, b):} Tensor network state energy $E$; the exact ground-state energy of $H_\text{HC}$ is $E_0 = 0$ by construction. \textbf{Panels~(c, d):} Relative error in the number of Hamiltonian cycles, measured as an observable and compared with exact values from Refs.~\cite{stoyan1996enumeration, Jacobsen_2007}. \textbf{Panels~(e, f):} Probability of sampling non-Hamiltonian multiloops (from 1000 samples). \textbf{Panels~(g, h):} Expected number of cycles in sampled configurations representing 2-factors (from 1000 samples).
    }
    \label{fig:convergence}
\end{figure*}

\begin{figure}
    \centering
    \includegraphics[width=0.7\linewidth]{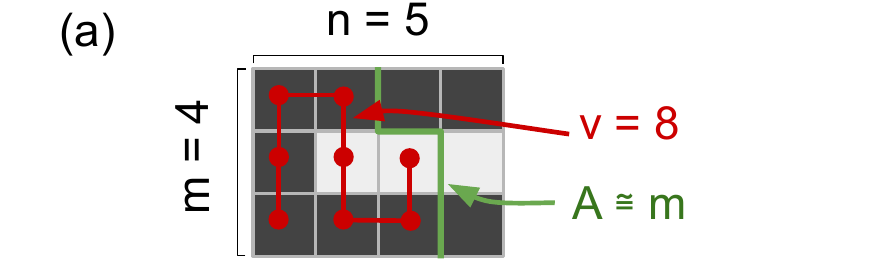}
    \includegraphics[width=\linewidth]{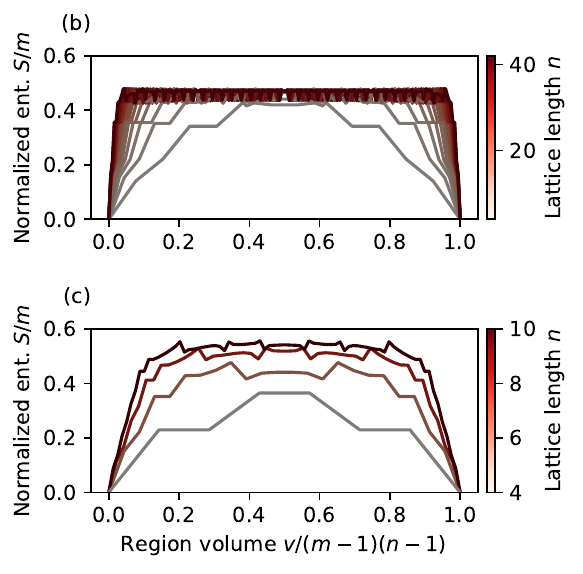}
    \caption{\textit{Normalized entanglement entropy $S$ calculated over bipartitions of a rectangular lattice.} \textbf{Panel~(a)}: the lattice shape is $m\times n$. The regions are defined by ordering the lattice sites along a snake path starting from the upper-left corner and selecting the first $v$ sites. The size of the boundary between the region and the remainder of the system scales with the lattice width $m$.
    The entanglement entropy is normalized by $m$ and is shown as a function of the ratio between the region size $v$ and the dual lattice size $(m-1)\times (n-1)$, for lattices of different lengths $n$. \textbf{Panel~(b)} corresponds to rectangular lattices with $m = 6$. \textbf{Panel~(c)} corresponds to square lattices with $m = n$.
    }
    \label{fig:entanglement}
\end{figure}

\begin{figure}
    \centering
    \includegraphics[width=0.85\linewidth]{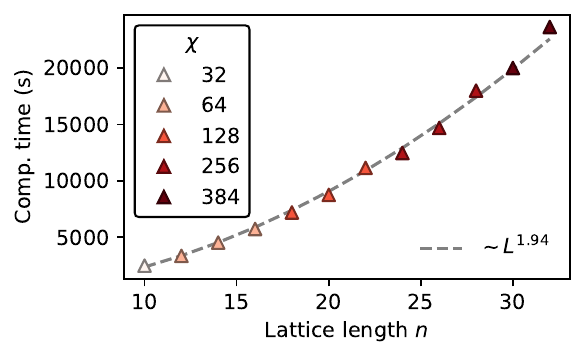}
    \caption{\textit{Computational time needed for approximating $\ket{X_\text{HC}}$ on a rectangular lattice, as a function of the lattice length.} We plot the time needed to construct an MPS approximation capable of encoding the total number of cycles with an error smaller than $0.5\%$. This takes into account both the required bond dimension (different colors of the markers) and the number of steps required for the DMRG algorithm to converge.
    }
    \label{fig:time_complexity}
\end{figure}

In this section, we employ tensor network (TN) methods~\cite{PhysRevB.94.165116, RevModPhys.77.259, SCHOLLWOCK201196, Montangero2018, Silvi2019} to approximate the ground state  of $\hat H_\text{HC}$ on a dual lattice with  $(m+1)\times( n+1)$ plaquettes, corresponding to an $m\times n$ rectangular lattice.

We encode the state $\ket{X_\text{HC}}$ as a \textit{Matrix Product State} (MPS), a one-dimensional array of rank-3 tensors connected through contracted indices. Fig.~\ref{fig:zigzag} illustrates the mapping from the dual lattice to the MPS. Each tensor has at most a shape $(\chi, 2, \chi)$, where $2$ is the dimension of the local Hilbert space and $\chi$ is the dimension of the contracted indices. The parameter $\chi$, called the \textit{bond dimension}, determines the accuracy of the MPS approximation. The 2D dual lattice is mapped to a 1D chain using a snake space-filling curve~\cite{Cataldi_2021}. Since all boundary plaquette states are fixed to $\sigma_{ij} = 0$, only the bulk $(n-1)\times(m-1)$ plaquettes need to be encoded in the MPS. The snake curve, therefore, starts from the top-left corner of the bulk, proceeds first along its shorter side, and covers all these sites, see Fig.~\ref{fig:zigzag}.

We obtain the MPS representation of $\ket{X_\text{HC}}$ using the density matrix renormalization group (DMRG) algorithm~\cite{RevModPhys.77.259}, as implemented in the tensor-network emulator suite \textit{Quantum TEA Leaves}~\cite{qtealeaves}. We consider rectangular lattices of shape $6\times n$ with $n$ in the range $[4,40]$, and square lattices of shape $n\times n$ with even $n$ in the range $[4,10]$. For each lattice shape, we perform DMRG calculations with bond dimension $\chi\in[32, 64, 128, 256, 384]$.

This TN framework provides a compressed encoding of the entire ensemble of Hamiltonian cycles. Instead of explicitly storing all configurations, which would require exponential memory, the MPS represents the coherent quantum sample $\ket{X_\text{HC}}$ with a tractable bond dimension. This compact representation enables the efficient calculation of expectation values, and supports efficient sampling of Hamiltonian cycles~\cite{Ballarin2024}, as well as their approximate counting. Indeed, as shown in Appendix~\ref{app:counting}, the total number of Hamiltonian cycles is given by the expectation value
\begin{equation}\label{eq:counting_exp_val}
\bra{X_\text{HC}}\bigotimes_{\substack{2\leq i\leq m\\2\leq j\leq n}}\hat P^x_{ij}\ket{X_\text{HC}}=\frac{|X_\text{HC}|}{2^{(m-1)(n-1)}}\;,
\end{equation}
where $P^x_i$ projects onto the positive eigenstate of the Pauli-$X$ operator. Sampling the expectation value in Eq.~\ref{eq:counting_exp_val} | either through classical Monte-Carlo methods or by measurements on a quantum computer | is extremely difficult due to the exponentially small fraction of Hamiltonian cycles.

In Fig.~\ref{fig:convergence}, we assess the quality of the MPS approximation of $\ket{X_\text{HC}}$ using several figures of merit for rectangular and square lattice geometries and different bond dimensions. Panels~(a) and~(b) show the energy of the MPS, i.e., the expectation value of the Hamiltonian $\hat{H}_\text{HC}$. Any energy larger than zero signals constraint violations as well as non-uniformity in the probability distribution encoded in the MPS. Panels~(c) and~(d) display the relative error
\begin{equation}
\varepsilon_{\#_\text{cycles}}=\frac{||X_\text{HC}|^\chi-|X_\text{HC}|^\text{exact}|}{|X_\text{HC}|^\text{exact}}
\end{equation}
where $|X_\text{HC}|^\text{exact}$ denotes the number of Hamiltonian cycles obtained via exact counting~\cite{stoyan1996enumeration,Jacobsen_2007}, and $|X_\text{HC}|^\chi$ derives from Eq.~\ref{eq:counting_exp_val}. A small relative error indicates that the MPS successfully captures the full ensemble of Hamiltonian cycles. The exact and MPS-based counting results are also compared in Tables~\ref{tab:counting_rect} and~\ref{tab:counting_square}, corresponding to rectangular and square lattices, respectively.

Panels~(e) and~(f) of Figure~\ref{fig:convergence} report the probability of obtaining multiloop configurations that violate the Hamiltonian condition, estimated from $1000$ samples drawn from the MPS. Finally, Panels~(h) and~(i) present the expectation value of the number of cycles obtained from the sampled 2-factor configurations. Since a Hamiltonian cycle contains exactly one loop, an exact MPS representation of $\ket{X_{\mathrm{HC}}}$ yields an expected cycle count of 1.  Deviations above 1, therefore mark contributions from multiloop 2-factors. Together, these last two measures quantify the quality of the MPS representation of the Hamiltonian-cycle ensemble.

As the bond dimension increases, all quality indicators in Figure~\ref{fig:convergence} improve. For a fixed bond dimension and rectangular lattice, the approximation errors increase less than polynomially as the system size grows and tend towards a stable value. In contrast, for square lattices, the quality deteriorates rapidly with increasing system size and fixed bond dimension. The bond-dimension dependence of the approximation quality for fixed rectangular shapes is discussed in Appendix~\ref{app:convergence}, where we observe that both the energy and the relative error in counting Hamiltonian cycles converge polynomially with the bond dimension, with the convergence rate decreasing as the system size increases. For square lattices instead, convergence only begins once the bond dimension exceeds a threshold value that grows rapidly with system size, preventing an efficient approximation.

The behavior of the quality factors in Fig.~\ref{fig:convergence} reflects the entanglement structure of the state $\ket{X_\text{HC}}$. In Fig.~\ref{fig:entanglement}, we plot the entanglement entropy of this state for various bipartitions along the snake-like ordering used to map the two-dimensional lattice onto a one-dimensional MPS. The entanglement entropy is normalized by the smaller lattice dimension $m$, which corresponds to the area of the bipartition boundary (see Panel~(a)). Panel~(b) shows results for lattices of fixed width $m=6$ and lengths $n \in \{4,\dots,10\}$, while Panel~(c) displays data for square lattices with $n=m=L$ and $L \in \{4,6,8,10\}$. In both cases, as the subsystem volume increases,, the size of the boundary associated with the lattice bipartition remains fixed. The normalized entanglement $S/m$ is nearly independent of both the subsystem volume $v$ and the boundary size $m$, consistent with an area-law scaling of the entanglement entropy. Appendix~\ref{app:convergence} further analyzes the dependence of the maximum bipartition entanglement entropy on the bond dimension, confirming these indications of area-law scaling.

This area-law scaling admits a simple geometric interpretation. The distribution over Hamiltonian cycles on the full lattice differs from the product of the distributions restricted to two disjoint regions because any Hamiltonian cycle necessarily crosses the interface between them, thereby inducing correlations. The strength of these correlations is controlled by the number of boundary crossings, which is upper-bounded by the boundary area.

The observed area-law entanglement implies that the bond dimension required for an accurate MPS approximation is independent of the lattice’s longer side and grows exponentially only with its shorter dimension. Consequently, the computational cost of constructing and manipulating the MPS is exponential in the smaller lattice dimension. Differently, as shown in Figure~\ref{fig:time_complexity}, when this dimension is fixed, we can approximate the entire Hamiltonian cycle ensemble in time that scales quadratically with the lattice length. Notably, the resulting MPS representation provides a compressed representation of the entire ensemble of Hamiltonian cycles. In this case, we are able to encode up to approximately $3.77\times 10^{28}$ Hamiltonian cycles within an MPS requiring only 380~MB of memory.

\begin{table}
\centering
\begin{tabular}{||c|c|c|c||}
\hline
$m \times n$ & $|X_\text{HC}|^{\chi=384}$ & $|X_\textbf{HC}|^\text{exact}$ & $\varepsilon_{\#_\text{cycles}}$\\
\hline
6$\times$4 & 37 & 37 & 0.0 \\
6$\times$6 & 1072 & 1072 & 0.0 \\
6$\times$8 & 32675 & 32675 & 0.0 \\
6$\times$10 & 1024028 & 1024028 & 0.0 \\
6$\times$12 & 3246380\textbf{5.26} & 32463802 & 1e-07 \\
6$\times$14 & 103391\textbf{8048} & 1033917350 & 6.75e-07 \\
6$\times$16 & 3.2989\textbf{15608}e+10 & 3.298906816e+10 & 2.67e-06 \\
6$\times$18 & 1.0533\textbf{52975}e+12 & 1.053349394e+12 & 3.4e-06 \\
6$\times$20 & 3.3643\textbf{84982}e+13 & 3.364354121e+13 & 9.17e-06 \\
6$\times$22 & 1.074\textbf{721981}e+15 & 1.074685815e+15 & 3.37e-05 \\
6$\times$24 & 3.433\textbf{550978}e+16 & 3.433060709e+16 & 0.000143 \\
6$\times$26 & 1.09\textbf{7166749}e+18 & 1.096704136e+18 & 0.000422 \\
6$\times$28 & 3.50\textbf{6935385}e+19 & 3.50348837e+19 & 0.000983 \\
6$\times$30 & 1.1\textbf{21357798}e+21 & 1.119214053e+21 & 0.00191 \\
6$\times$32 & 3.5\textbf{87177531}e+22 & 3.575412358e+22 & 0.00328 \\
6$\times$34 & 1.14\textbf{7992622}e+24 & 1.142192547e+24 & 0.00505 \\
6$\times$36 & 3.6\textbf{75615823}e+25 & 3.648821244e+25 & 0.00729 \\
6$\times$38 & 1.1\textbf{77146356}e+27 & 1.165643875e+27 & 0.00977 \\
6$\times$40 & 3.7\textbf{71580754}e+28 & 3.723738685e+28 & 0.0127 \\
\hline
\end{tabular}
\caption{Number of Hamiltonian cycles for different shapes of a rectangular lattice. Comparison between exact counting $|X_\text{HC}|^\text{exact}$~\cite{stoyan1996enumeration,Jacobsen_2007} and expectation value $|X_\text{HC}|^{\chi=384}=\bra{X_\text{HC}}\bigotimes_{\substack{2\leq i\leq m\\2\leq j\leq n}}2\hat P^x_{ij}\ket{X_\text{HC}}$ for bond dimension $\chi=384$. Digits that differ from the exact values are highlighted in bold. $\varepsilon_{\#_\text{cycles}}$ is the relative error of the expectation value with respect to the exact counting, whose scaling is depicted in Figure~\ref{fig:convergence} Panel~(c).}
\label{tab:counting_rect}
\end{table}

\begin{table}
\centering
\begin{tabular}{||c|c|c|c||}
\hline
$m\times n$ & $|X_\textbf{HC}|^{\chi=384}$ & $|X_\textbf{HC}|^\text{exact}$ & $\varepsilon_{\#_\text{cycles}}$\\
\hline
4$\times$4 & 6 & 6 & 0.0 \\
6$\times$6 & 1072 & 1072 & 0.0 \\
8$\times$8 & 4638576.\textbf{113} & 4638576 & 2.44e-08 \\
10$\times$10 & 4.\textbf{842640256}e+11 & 4.672604566e+11 & 0.0351 \\
\hline
\end{tabular}
\caption{Number of Hamiltonian cycles for different sizes of a square lattice. Comparison between exact counting $|X_\text{HC}|^\text{exact}$~\cite{stoyan1996enumeration,Jacobsen_2007} and expectation value $|X_\text{HC}|^{\chi=384}=\bra{X_\text{HC}}\bigotimes_{\substack{2\leq i\leq m\\2\leq j\leq n}}2\hat P^x_{ij}\ket{X_\text{HC}}$ for bond dimension $\chi=384$. Digits that differ from the exact values are highlighted in bold. $\varepsilon_{\#_\text{cycles}}$ is the relative error of the expectation value with respect to the exact counting, whose scaling is depicted in Figure~\ref{fig:convergence} Panel~(d).}
\label{tab:counting_square}
\end{table}

\section{Conclusions and outlook}\label{sec:conclusions}

We have constructed a local, frustration-free Hamiltonian whose unique ground state is a coherent quantum sample encoding the complete set of Hamiltonian cycles on a rectangular lattice. The construction relies on local constraints enforcing the Hamiltonian condition together with a family of topology-preserving local transformation rules. These ingredients make it possible to encode a highly non-local combinatorial property—the existence of a Hamiltonian cycle—into the ground state of a local quantum Hamiltonian.

This approach improves on formulations based on classical Hamiltonians, where enforcing global topological constraints requires an exponentially large number of non-local terms. By encoding the desired ensemble in a coherent quantum superposition rather than in a degenerate manifold of classical configurations, this strategy enables a quadratic speedup over classical Monte Carlo through the application of amplitude amplification on the ensemble of Hamiltonian cycles~\cite{Brassard, Grover_1998}. We have further shown how to extend this strategy to encode Boltzmann distributions at finite temperature for both polymers and heteropolymers.

Restricting the analysis to two-dimensional closed paths simplified both the analytical construction of the Hamiltonian and the numerical approximation of the ground state. Nevertheless, the proposed framework applies to more general polymer ensembles, which can be represented by open Hamiltonian paths on a three-dimensional lattice. Specifically, the equational reasoning~\cite{quantum_eq_reasoning} approach to constructing the parent Hamiltonian is feasible whenever a complete and sound set of transformation rules exists to connect all and only the feasible polymer configurations.  In the Monte Carlo literature, such a set is employed to explore polymer configurations ergodically~\cite {Mansfield_sampling}. Any relevant speedup in the computation of the partition function for a given heteropolymer sequence may prove invaluable in the context of protein design~\cite{PRXLife.2.043012}.

We have also shown that MPS provide accurate approximations of the entire ensemble of Hamiltonian cycles when one lattice dimension is fixed. This representation enables the approximate counting of Hamiltonian cycles through tensor-network contractions. In particular, the tensor-network approach allows the computation of expectation values that are inaccessible to classical Monte-Carlo, such as the probability of encountering a specific configuration. As a result, tensor-network methods offer a polynomial-time alternative for quasi–1D geometries and remain competitive for moderately sized two-dimensional lattices.

The tensor network method introduced in this study represents a generalization of transfer matrix methods. While transfer matrix methods are considered state-of-the-art for exact counting~\cite{stoyan1996enumeration, Jacobsen_2007}, matrix product states enable the approximation of expectation values for arbitrary product operators. Similar to transfer matrix methods, these expectation values are computed by multiplying transfer matrices, which are formed by contracting each local tensor in the matrix product state with the corresponding local observable. This analogy also applies to computational time: due to area-law entanglement, both the proposed approach and standard transfer matrix methods demonstrate an exponential increase in transfer matrix dimension and runtime relative to the shorter dimension of the lattice. In contrast to transfer matrix methods, which construct the transfer matrix based on combinatorial principles, the present approach utilizes density matrix renormalization to economize computational resources, accepting a less accurate approximation, for example, by employing a small bond dimension.

In the future, a comprehensive understanding of the complexity of ground-state preparation will be essential for evaluating the full extent of quantum advantage. While we did not assess particular preparation strategies in this work, we note that quantum walk dynamics~\cite{Venegas-Andraca2012} is a good candidate for this purpose, since it can achieve a quadratic speedup in generating coherent quantum samples compared to classical Monte Carlo methods~\cite{doi:10.1098/rspa.2015.0301}.

\begin{acknowledgments}
We acknowledge M. Baiesi, M. Ballarin, M. Rigobello, M. Tesoro, 
A. Braghetto, E. Orlandini, and G. Cataldi for useful discussions. The research leading to these results has received funding from the following organizations: European Union via Italian Research Center on HPC, Big Data and Quantum Computing (NextGenerationEU Project No. CN00000013), project EuRyQa (Horizon 2020), project PASQuanS2 (Quantum Technology Flagship); Italian Ministry of University and Research (MUR) via: Quantum Frontiers (the Departments of Excellence 2023-2027); the World Class Research Infrastructure - Quantum Computing and Simulation Center (QCSC) of Padova University; Istituto Nazionale di Fisica Nucleare (INFN): iniziativa specifica IS-QUANTUM; the German Federal Ministry of Education and Research (BMBF) via the project QRydDemo. We acknowledge computational resources from Cloud Veneto, as well as computational time on Cineca’s Leonardo machine.
\end{acknowledgments}

\appendix

\section{Operators encoding the action of rules in $\mathcal{E}$}\label{app:rules}

Here we list the operators encoding the action of rules in $\mathcal{E}$. These operators are:
\begin{align}
E_1\rightarrow\hat e^{[1]}_{ij} = & \widehat{P}^{[1]}_{i,j+1}\widehat{P}^{[1]}_{i,j+2}\widehat{P}^{[0]}_{i+1,j}\hat\sigma^+_{i+1,j+1}\hat\sigma^-_{i+1,j+2}\widehat{P}^{[0]}_{i+1,j+3}\nonumber\\&\otimes\widehat{P}^{[1]}_{i+2,j+1}\widehat{P}^{[1]}_{i+2,j+2}\nonumber\\
E_2\rightarrow\hat e^{[2]}_{ij} =& \mathcal{R}\left[e^{[1]}_{ij}\right] \nonumber \\
E_3\rightarrow\hat e^{[3]}_{ij}  =& \widehat{P}^{[0]}_{i,j+1}\widehat{P}^{[0]}_{i,j+2}\widehat{P}^{[1]}_{i+1,j}\hat\sigma^+_{i+1,j+1}\hat\sigma^-_{i+1,j+2}\widehat{P}^{[1]}_{i+1,j+3}\nonumber\\&\otimes\widehat{P}^{[0]}_{i+2,j+1}\widehat{P}^{[0]}_{i+2,j+2}\nonumber\\
E_4\rightarrow\hat e^{[4]}_{ij} =& \mathcal{R}\left[e^{[3]}_{ij}\right] \nonumber \\
E_5\rightarrow\hat e^{[5]}_{ij} =& \widehat{P}^{[0]}_{i,j+1}\widehat{P}^{[1]}_{i+1,j}\hat\sigma^+_{i+1,j+1}\widehat{P}^{[1]}_{i+1,j+2}\widehat{P}^{[1]}_{i+2,j}\widehat{P}^{[0]}_{i+2,j+1}\nonumber\\&\otimes\hat\sigma^-_{i+2,j+2}\widehat{P}^{[0]}_{i+2,j+3}\widehat{P}^{[1]}_{i+3,j+1}\widehat{P}^{[1]}_{i+3,j+2}\nonumber\\
E_6\rightarrow\hat e^{[6]}_{ij} =& \mathcal{R}\left[e^{[5]}_{ij}\right] \nonumber \\
E_7\rightarrow\hat e^{[7]}_{ij} =& \mathcal{R}\left[e^{[6]}_{ij}\right]\nonumber \\ E_8\rightarrow\hat e^{[8]}_{ij} =& \mathcal{R}\left[e^{[7]}_{ij}\right] \nonumber \\ E_9\rightarrow
\hat e^{[9]}_{ij} =& \widehat{P}^{[1]}_{i,j+1}\widehat{P}^{[0]}_{i+1,j}\hat\sigma^+_{i+1,j+1}\widehat{P}^{[0]}_{i+1,j+2}\widehat{P}^{[0]}_{i+2,j}\widehat{P}^{[1]}_{i+2,j+1}\nonumber\\&\otimes\hat\sigma^-_{i+2,j+2}\widehat{P}^{[1]}_{i+2,j+3}\widehat{P}^{[0]}_{i+3,j+1}\widehat{P}^{[0]}_{i+3,j+2}\nonumber \\
E_{10}\rightarrow\hat e^{[10]}_{ij} =& \mathcal{R}\left[e^{[9]}_{ij}\right] \nonumber \\ E_{11}\rightarrow\hat e^{[11]}_{ij} =& \mathcal{R}\left[e^{[10]}_{ij}\right] \nonumber \\ E_{12}\rightarrow\hat e^{[12]}_{ij}=& \mathcal{R}\left[e^{[11]}_{ij}\right]\;,
\end{align}
where $\widehat{P}^{[0]}_{ij} = \left(\ket{0}\bra{0}\right)_{ij}$, $\widehat{P}^{[1]]}_{ij} = \left(\ket{1}\bra{1}\right)_{ij}$, $\hat\sigma^{+}_{ij} = \left(\ket{1}\bra{0}\right)_{ij}$, and $\hat\sigma^{-}_{ij} = \left(\ket{0}\bra{1}\right)_{ij}$. The super-operator $\mathcal{R}$ rotates the region of the dual lattice on which a rule acts by $\pi/2$.

\section{First energy gap of $\hat H_\text{HC}$}\label{app:energy_gap}

In this appendix, we study the scaling of the first non-zero eigenvalue $\Delta_{\alpha}$ of the Hamiltonian $\hat H_\text{HC}$ restricted to the orthogonal subspace $\mathcal{X}_\alpha$.

\subsection{Halmos's two-spaces theorem and the first energy gap}

From Eq.~\ref{eq:hhc_decomposition}, the Hamiltonian $\hat H_\text{HC}$ restricted to $\mathcal{X}_\alpha$ can be written as
\begin{equation}
    \hat H_\text{HC}\Big|_{\mathcal{X}_\alpha} = \left(\hat L_\text{E} + \hat H_\text{L}\right)\Big|_{\mathcal{X}_\alpha}\;.
\end{equation}

Let $L_\alpha$ ($L_\alpha^\perp$) and $N_\alpha$ ($N_\alpha^\perp$) denote, respectively, the images (kernels) of $\hat L_\text{E}$ and $\hat H_\text{L}$ restricted to $\mathcal{X}_\alpha$:
\begin{align}
    L_\alpha &:= \ker\left(\hat L_\text{E}\right)^\perp \cap \mathcal{X}_\alpha\nonumber\\
    L_\alpha^\perp &:= \ker\left(\hat L_\text{E}\right) \cap \mathcal{X}_\alpha\nonumber\\
    N_\alpha &:= \ker\left(\hat H_\text{L}\right)^\perp \cap \mathcal{X}_\alpha\nonumber\\
    N_\alpha^\perp &:= \ker\left(\hat H_\text{L}\right) \cap \mathcal{X}_\alpha\;,
\end{align}
where, with a slight abuse of notation, the orthogonal complements on the left-hand side are taken within $\mathcal{X}_\alpha$, while those on the right-hand side are taken in the full Hilbert space $\mathcal{H}$.  

By construction (see main text):
\begin{itemize}
    \item $L_\alpha^\perp$ is spanned by the equal-weight superposition $\ket{X_\alpha}$ of all computational basis states encoding elements of $X_\alpha$;
    \item $N_\alpha^\perp$ is spanned by all computational basis states encoding elements of $X_\alpha$ that contain no local loops.
\end{itemize}

$\ket{X_\alpha}$ overlaps configurations both with and without local cycles. Hence, it is not contained in $N_\alpha^\perp$, and
\begin{equation}\label{eq:empty_intersection}
    L_\alpha^\perp\cap N_\alpha^\perp = \varnothing\;.
\end{equation}

Let $\hat P_\alpha$ and $\hat Q_\alpha$ denote the projectors from $\mathcal{X}_\alpha$ onto the subspaces $L_\alpha$ and $N_\alpha$, respectively; and let $\Delta_{\mathcal{X}_\alpha}$ and $\Delta_{\text{L},\alpha}$ respectively be the smallest non-zero eigenvalues of $\hat L_\text{E}|_{\mathcal{X}_\alpha}$ and $\hat H_\text{L}|_{\mathcal{X}_\alpha}$. We then have
\begin{align}\label{eq:projectors_bound}
     \bra{\psi} \hat L_\text{E} + \hat H_\text{L}\ket{\psi} &= \bra{\psi} \hat L_\text{E} \ket{\psi} + \bra{\psi} \hat H_\text{L}\ket{\psi}\nonumber\\
     &\geq  \bra{\psi} \Delta_{\mathcal{X}_\alpha}\hat P_\alpha \ket{\psi} + \bra{\psi} \Delta_{\text{L},\alpha}\hat Q_\alpha \ket{\psi}\nonumber\\
     &= \bra{\psi} \big(\Delta_{\mathcal{X}_\alpha}\hat P_\alpha + \Delta_{\text{L},\alpha}\hat Q_\alpha\big) \ket{\psi}
\end{align}
for all $\ket{\psi}\in \mathcal{X}_\alpha$, so that the smallest eigenvalue of $\left(\hat L_\text{E} + \hat H_\text{L}\right)\Big|_{\mathcal{X}_\alpha}$ is bounded below by the smallest eigenvalue $\lambda_\alpha$ of $\Delta_{\mathcal{X}_\alpha}\hat P_\alpha + \Delta_{\text{L},\alpha}\hat Q_\alpha$:
\begin{align}\label{eq:delta_vs_lambda}
    \Delta_{\alpha} &= \min_{\ket{\psi}\in \mathcal{X}_\alpha}\frac{\bra{\psi}  \hat L_\text{E} + \hat H_\text{L}\ket{\psi}}{\langle\psi|\psi\rangle}\nonumber\\
    &\geq \min_{\ket{\psi}\in \mathcal{X}_\alpha}\frac{\bra{\psi}  \big(\Delta_{\mathcal{X}_\alpha}\hat P_\alpha + \Delta_{\text{L},\alpha}\hat Q_\alpha\big)\ket{\psi}}{\langle\psi|\psi\rangle} := \lambda_\alpha\;.
\end{align}

The spectral structure of the operator $\Delta_{\mathcal{X}_\alpha}\hat P_\alpha + \Delta_{\text{L},\alpha}\hat Q_\alpha$ can be analyzed using Halmos's two-subspaces theorem.

\begin{theorem}\label{th:halmos}
    \textbf{Halmos's two-subspaces theorem~\cite{halmos1969two, BOTTCHER20101412}:}  
    Let $\hat P_\alpha$ and $\hat Q_\alpha$ be the projectors onto subspaces $L_\alpha$ and $N_\alpha$, respectively. Then, the Hilbert space $\mathcal{X}_\alpha$ can be decomposed as
\begin{equation}
    \mathcal{X}_\alpha = L_\alpha\cap N_\alpha \oplus L_\alpha\cap N_\alpha^\perp \oplus L_\alpha^\perp \cap N_\alpha \oplus L_\alpha^\perp\cap N_\alpha^\perp \oplus M_i
\end{equation}
and $\hat P_\alpha$ and $\hat Q_\alpha$ are block-diagonal in this decomposition. In particular, they can be written as
\begin{equation}
    \hat P_\alpha = \id\oplus\id\oplus 0 \oplus 0 \oplus 
\begin{pmatrix}
  \id &   0   \\
  0 &   0   
\end{pmatrix},
\end{equation}
\begin{equation}
    \hat Q_\alpha = \id\oplus 0 \oplus \id \oplus 0 \oplus 
\begin{pmatrix}
  \id-S_\alpha^2 &   S_\alpha\sqrt{\id-S_\alpha^2}   \\
  S_\alpha\sqrt{\id-S_\alpha^2} &   S_\alpha^2   
\end{pmatrix},
\end{equation}
for some Hermitian operator $S_\alpha^2$ satisfying
\begin{equation}\label{eq:s_bound}
    0\leq S_\alpha^2 \leq \id\;.
\end{equation}
\end{theorem}

Since $L_\alpha^\perp\cap N_\alpha^\perp = \varnothing$, this component can be omitted from the decomposition. The Halmos' theorem then yields the following block-diagonal form for $\Delta_{\mathcal{X}_\alpha}\hat P_\alpha + \Delta_{\text{L},\alpha}\hat Q_\alpha$:
\begin{align}\label{eq:halmos_decomposition}
    &\Delta_{\mathcal{X}_\alpha}\hat P_\alpha + \Delta_{\text{L},\alpha}\hat Q_\alpha = (\Delta_{\mathcal{X}_\alpha}+\Delta_{\text{L},\alpha})\id\oplus\Delta_{\mathcal{X}_\alpha}\id\nonumber\\
    &\oplus \Delta_{\text{L},\alpha}\id\nonumber\\
    &\oplus  
\begin{pmatrix}
  (\Delta_{\mathcal{X}_\alpha} + \Delta_{\text{L},\alpha})\id-\Delta_{\text{L},\alpha}S_\alpha^2 &   \Delta_{\text{L},\alpha}S_\alpha\sqrt{\id-S_\alpha^2}   \\
  \Delta_{\text{L},\alpha}S_\alpha\sqrt{\id-S_\alpha^2} &   \Delta_{\text{L},\alpha}S_\alpha^2   
\end{pmatrix}.
\end{align}

The eigenvalues $\{x\}$ of $\Delta_{\mathcal{X}_\alpha}\hat P_\alpha + \Delta_{\text{L},\alpha}\hat Q_\alpha$ can be found by imposing the eigenvalues equation on the Halmos' decomposition in Eq.\ref{eq:halmos_decomposition}:
\begin{align}
    &\det\left[\Delta_{\mathcal{X}_\alpha}\hat P_\alpha + \Delta_{\text{L},\alpha}\hat Q_\alpha - x\id\right]=\nonumber\\
    &\left(\Delta_{\mathcal{X}_\alpha}+\Delta_{\text{L},\alpha}-x\right)^{\dim\left(L_\alpha\cap N_\alpha\right)}\nonumber\\
    &\cdot\left(\Delta_{\mathcal{X}_\alpha}-x\right)^{\dim\left(L_\alpha\cap N_\alpha^\perp\right)}\cdot\left(\Delta_{\text{L},\alpha}-x\right)^{\dim\left(L_\alpha^\perp\cap N_\alpha\right)}\nonumber\\
   &\cdot\det\!\left(\!
    \begin{array}{ll}
  \left(\Delta_{\mathcal{X}_\alpha} + \Delta_{\text{L},\alpha} - x\right)\id-\Delta_{\text{L},\alpha}S_\alpha^2 &   \Delta_{\text{L},\alpha}S_\alpha\sqrt{\id-S_\alpha^2}   \\
  \Delta_{\text{L},\alpha}S_\alpha\sqrt{\id-S_\alpha^2} &   \Delta_{\text{L},\alpha}S_\alpha^2 - x\id \\
    \end{array}
    \right)\nonumber\\
    &=0\;.
\end{align}

The last determinant in the latter equation can be determined via the Schur complement formula, taking into account that the blocks of the matrix commute, so that we obtain

\begin{align}
    &\det\left(
    \begin{array}{ll}
  \left(\Delta_{\mathcal{X}_\alpha} + \Delta_{\text{L},\alpha} - x\right)\id-\Delta_{\text{L},\alpha}S_\alpha^2 &   \Delta_{\text{L},\alpha}S_\alpha\sqrt{\id-S_\alpha^2}   \\
  \Delta_{\text{L},\alpha}S_\alpha\sqrt{\id-S_\alpha^2} &   \Delta_{\text{L},\alpha}S_\alpha^2 - x\id \\
    \end{array}
    \right) \nonumber\\
    &=
    \det\Big(\left[\left(\Delta_{\mathcal{X}_\alpha} + \Delta_{\text{L},\alpha} - x\right)\id-\Delta_{\text{L},\alpha}S_\alpha^2\right] \left(\Delta_{\text{L},\alpha}S_\alpha^2 - x\id\right) -\nonumber\\
    &\left(\Delta_{\text{L},\alpha}S_\alpha\sqrt{\id-S_\alpha^2}\right)^2\Big)\nonumber\\
    & = \det\left(x^2-x(\Delta_{\mathcal{X}_\alpha}+\Delta_{\text{L},\alpha}) + \Delta_{\mathcal{X}_\alpha}\Delta_{\text{L},\alpha} S_\alpha^2\right)\nonumber\\
    & = \!\!\!\!\!\!\prod_{s_i\in \sigma(S_\alpha)}\!\left[x-\frac{\left(\Delta_{\mathcal{X}_\alpha}\!\!+\!\Delta_{\text{L},\alpha}\right) \pm \sqrt{\left(\Delta_{\mathcal{X}_\alpha}\!\!+\!\Delta_{\text{L},\alpha}\right)^2\!-4\Delta_{\mathcal{X}_\alpha}\Delta_{\text{L},\alpha}s^2}}{2}\right],
\end{align}
where $\sigma(S_\alpha)$ is the spectrum of $S_\alpha$.

The smallest eigenvalue is thus
\begin{align}\label{eq:min_a_b_sqrt}
    \lambda_\alpha &= \min\Big(\Delta_{\mathcal{X}_\alpha},\Delta_{\text{L},\alpha},\nonumber\\
    &\frac{\left(\Delta_{\mathcal{X}_\alpha}+\Delta_{\text{L},\alpha}\right) - \sqrt{\left(\Delta_{\mathcal{X}_\alpha}+\Delta_{\text{L},\alpha}\right)^2-4\Delta_{\mathcal{X}_\alpha}\Delta_{\text{L},\alpha}\|S_\alpha^2\|}}{2}\Bigg).
\end{align}

Taking into account Eq.~\ref{eq:s_bound} we can write
\begin{align}
    &\left(\Delta_{\mathcal{X}_\alpha}+\Delta_{\text{L},\alpha}\right) - \sqrt{\left(\Delta_{\mathcal{X}_\alpha}+\Delta_{\text{L},\alpha}\right)^2-4\Delta_{\mathcal{X}_\alpha}\Delta_{\text{L},\alpha}\|S_\alpha^2\|}\nonumber\\
    &\leq \left(\Delta_{\mathcal{X}_\alpha}+\Delta_{\text{L},\alpha}\right) - \sqrt{\left(\Delta_{\mathcal{X}_\alpha}-\Delta_{\text{L},\alpha}\right)^2}\nonumber\\
    &= \min \big(2\Delta_{\mathcal{X}_\alpha}, 2\Delta_{\text{L},\alpha}\big)\;,
\end{align}
so that Eq.~\ref{eq:min_a_b_sqrt} reduces to
\begin{equation}\label{eq:second_bound}
    \lambda_\alpha = \frac{\left(\Delta_{\mathcal{X}_\alpha}+\Delta_{\text{L},\alpha}\right) - \sqrt{\left(\Delta_{\mathcal{X}_\alpha}+\Delta_{\text{L},\alpha}\right)^2-4\Delta_{\mathcal{X}_\alpha}\Delta_{\text{L},\alpha}\|S_\alpha^2\|}}{2}\;.
\end{equation}

To evaluate $\|S_\alpha^2\|$, consider the operator $\hat R_\alpha$, expressed in the Halmos decomposition as
\begin{equation}
    \hat R_\alpha = 0 \oplus 0 \oplus 0 \oplus 
\begin{pmatrix}
  0 & 0\\
  0 & \id
\end{pmatrix}.
\end{equation}
This operator projects onto $M \cap L^\perp$.  

Using the decomposition of $\hat Q_\alpha$, we have
\begin{equation}
    \hat R_\alpha \hat Q_\alpha \hat R_\alpha = 0 \oplus 0 \oplus 0 \oplus 
\begin{pmatrix}
  0 & 0\\
  0 & S_\alpha^2
\end{pmatrix}.
\end{equation}

The corresponding norm can be expressed in the bra–ket formalism. The subspace $M \cap L^\perp$ consists of states lying in $L^\perp$ that belong neither to $N$ nor to $N^\perp$. These correspond to equal-amplitude superpositions over sets of topologically equivalent 2-factors that can be contracted to local loops or expanded to non-local ones. The only state in $\mathcal{X}_\alpha$ satisfying this condition is $\ket{X_\alpha}$, so that $\hat R_\alpha=\ket{X_\alpha}\bra{X_\alpha}$ and
\begin{equation}
        \|S_\alpha^2\| = \| \hat R_\alpha \hat Q_\alpha\hat R_\alpha\| = \bra{X_\alpha}\hat Q_\alpha\ket{X_\alpha}\;.
\end{equation}

Substituting the latter equation into Eq.~\ref{eq:second_bound}, we obtain
\begin{align}\label{eq:lambda_i}
    \lambda_\alpha =& \frac{\Delta_{\mathcal{X}_\alpha}+\Delta_{\text{L},\alpha}}{2} \nonumber\\
    &- \sqrt{\left(\frac{\Delta_{\mathcal{X}_\alpha}+\Delta_{\text{L},\alpha}}{2}\right)^2-\Delta_{\mathcal{X}_\alpha}\Delta_{\text{L},\alpha}\cdot\bra{X_\alpha}\hat Q_\alpha\ket{X_\alpha}}\;.
\end{align}

\subsection{Lower bound for $\bra{X_\alpha}\hat Q_\alpha\ket{X_\alpha}$}

\begin{figure}
    \centering
    \includegraphics[width=\linewidth]{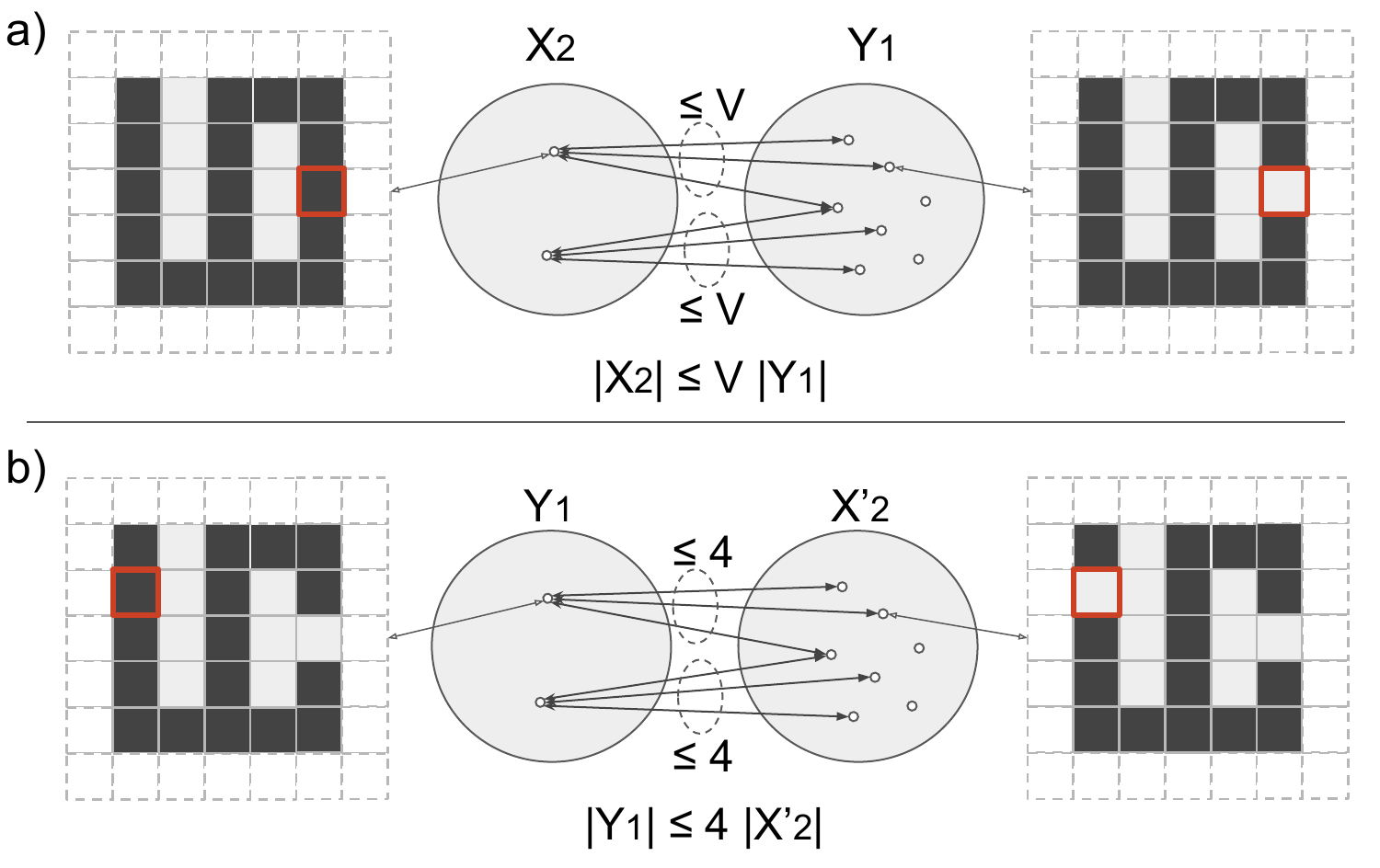}
    \caption{\textit{Relation between the size of the set $X_\alpha$ of topologically equivalent 2-factors and its subset $X_\alpha'$ containing configurations with at least one local loop.} The set $Y_{l_\alpha-1}$ denotes all possible 2-factors with $l_\alpha - 1$ cycles, where $l_\alpha$ is the number of cycles for the configurations in $X_\alpha$. \textbf{Panel~(a):} Configurations in $X_\alpha$ can be obtained by changing the color of a single plaquette of a configuration belonging to the set $Y_{l_\alpha-1}$. \textbf{Panel~(b):} Configurations in $Y_{l_\alpha-1}$ can be obtained by changing the color of a single plaquette in a configuration belonging to $X_\alpha'$, here the modified plaquette is adjacent to a local cycle.
}
    \label{fig:fraction_bound}
\end{figure}

We now aim to lower-bound the expectation value
\begin{equation}
    \bra{X_\alpha}\hat Q_\alpha\ket{X_\alpha}\,.
\end{equation}
Recall that $\hat Q_\alpha$ is a projector on the rank of $\hat H_L|_{\mathcal{X}_\alpha}$, so it assigns the eigenvalue $0$ to configurations that do not contain local loops, and $1$ to those that contain one or more local loops.  
Let $X_\alpha' \subset X_\alpha$ denote the subset of configurations in $X_\alpha$ that contain at least one local loop. Then,
\begin{equation}\label{eq:ratio_bound}
    \bra{X_\alpha}\hat Q_\alpha\ket{X_\alpha} =\!
    \sum_{\substack{\bm\sigma,\bm\sigma'\in X_\alpha\\\bm\sigma''\in X_\alpha'}}\frac{1}{|X_\alpha|}
    \bra{\bm\sigma}\bm\sigma''\rangle\langle\bm\sigma''\ket{\bm\sigma'}
    = \frac{|X_\alpha'|}{|X_\alpha|}\;,
\end{equation}
that is, $\bra{X_\alpha}\hat Q_\alpha\ket{X_\alpha}$ represents the fraction of topologically equivalent 2-factors that contain at least one local cycle.

We can bound this fraction using the following argument.  
Let $Y_{l_\alpha-1}$ be the set of all possible 2-factors containing $l_\alpha-1$ cycles, where $l_\alpha$ is the number of cycles of the configurations in $X_\alpha$.  
Every element of $X_\alpha$ can be obtained from some element of $Y_{l_\alpha-1}$ by changing the color of a single plaquette in the dual lattice (Fig.~\ref{fig:fraction_bound}a).  
If $V$ denotes the number of plaquettes in the lattice, this defines a mapping $Y_{l_\alpha-1} \leftrightarrow X_\alpha$ such that each configuration in $X_\alpha$ is connected to at least one and at most $V$ configurations in $Y_{l_\alpha-1}$.  
This implies
\begin{equation}\label{eq:set_bound1}
    |X_\alpha| \leq |Y_{l_\alpha-1}| \cdot V\,.
\end{equation}

Analogously, each configuration in $Y_{l_\alpha-1}$ can be generated from one configuration in $X_\alpha'$ by merging a local loop with one of its adjacent loops.  
This operation is equivalent to changing the color of a single plaquette in the dual lattice, as discussed in the main text.  
In this case, the plaquette must be adjacent to a local loop (Fig.~\ref{fig:fraction_bound}b).  
Since each local loop has at most four adjacent plaquettes, this defines a mapping $X_\alpha' \leftrightarrow Y_{l_\alpha-1}$ such that each configuration in $Y_{l_\alpha-1}$ is connected to at least one and at most four configurations in $X_\alpha'$.  
Hence,
\begin{equation}\label{eq:set_bound2}
    |Y_{l_\alpha-1}| \leq 4\,|X_\alpha'|\,.
\end{equation}

Combining Eqs.~\ref{eq:set_bound1} and~\ref{eq:set_bound2}, we find
\begin{equation}
   \bra{X_\alpha}\hat Q_\alpha\ket{X_\alpha}= \frac{|X_\alpha'|}{|X_\alpha|}
    \geq
    \frac{|Y_{l_\alpha-1}|}{|Y_{l_\alpha-1}|\cdot 4V}
    = \frac{1}{4V}\,.
\end{equation}

Substituting this bound into Eq.~\ref{eq:lambda_i} yields
\begin{align}
    \lambda_\alpha
    \geq& \frac{1}{2}\Big(
        \Delta_{\mathcal{X}_\alpha}
        + \Delta_{\text{L},\alpha}
        - \sqrt{
            \left(\Delta_{\mathcal{X}_\alpha}
            + \Delta_{\text{L},\alpha}\right)^2
            - \frac{
                \Delta_{\mathcal{X}_\alpha}
                \Delta_{\text{L},\alpha}}{V}}
    \Big)\,.
\end{align}

This bound can be simplified by considering that $\Delta_{\text{L},\alpha} = 1$. Then, in the thermodynamic limit $1/V \ll 1$, we get
\begin{equation}
    \lambda_\alpha \geq \frac{\Delta_{\mathcal{X}_\alpha}}{4V(1+\Delta_{\mathcal{X}_\alpha})}\,.
\end{equation}

Replacing the latter inequality in Eq.~\ref{eq:delta_vs_lambda} we finally obtain:
\begin{equation}
    \Delta_\alpha \geq \frac{\Delta_{\mathcal{X}_\alpha}}{4V(1+\Delta_{\mathcal{X}_\alpha})}\,.
\end{equation}

\section{Partition function as an observable}\label{app:counting}

Here, we show how the total number of Hamiltonian cycles, that is, the infinite temperature partition function, is related to the expectation value of
\begin{equation}
\bigotimes_{\substack{2\leq i\leq m\\2\leq j\leq n}}\hat P^x_{ij}\;,
\end{equation}
for a system in the state $\ket{X_\text{HC}}$.

Explicitly, the expectation value is
\begin{align}
&\bra{X_\text{HC}}\bigotimes_{\substack{2\leq i\leq m\\2\leq j\leq n}}\hat P^x_{ij}\ket{X_\text{HC}}\nonumber\\
&=\frac{1}{|X_\text{HC}|}\sum_{\bm\sigma,\bm\sigma'\in X_\text{HC}}\bra{\bm\sigma'}\bigotimes_{\substack{2\leq i\leq m\\2\leq j\leq n}}\hat P^x_{ij}\ket{\bm\sigma}\;.
\end{align}

The computational basis states are tensor products of local configurations, i.e., $\ket{\bm\sigma}=\bigotimes_{ij}\ket{\sigma_{ij}}$. Moreover, the plaquettes on the boundary of the dual lattice are fixed in the state $\ket{\sigma_{ij}}=\ket{0}$, so that $\langle \sigma'_{ij}|\sigma_{ij}\rangle=1$ for all $(i,j)\notin[2,\dots,m-1]\times[2,\dots,n-1]$. Thus, the previous equation becomes

\begin{align}
&\bra{X_\text{HC}}\bigotimes_{\substack{2\leq i\leq m\\2\leq j\leq n}}\hat P^x_{ij}\ket{X_\text{HC}}\nonumber\\
&=\frac{1}{|X_\text{HC}|}\sum_{\bm\sigma,\bm\sigma'\in X_\text{HC}}\prod_{\substack{2\leq i\leq m\\2\leq j\leq n}}\bra{\sigma'_{ij}}\hat P^x_{ij}\ket{\sigma_{ij}}\prod_{\substack{(i,j)\notin \\ [2,\dots,m-1]\times \\ [2,\dots,n-1]}}\langle0\ket{0}\nonumber\\
&=\frac{1}{|X_\text{HC}|}\sum_{\bm\sigma,\bm\sigma'\in X_\text{HC}}\prod_{\substack{2\leq i\leq m\\2\leq j\leq n}}\frac{1}{2}\nonumber\\
&=\frac{|X_\text{HC}|}{2^{(m-1)(n-1)}}\;,
\end{align}
which allows us to infer the size $|X_\text{HC}|$ of the set of Hamiltonian cycles from the expectation value of $\bigotimes_{\substack{2\leq i\leq m\\2\leq j\leq n}}\hat P^x_{ij}$.

\section{Convergence in bond dimension}\label{app:convergence}

\begin{figure*}[t]
    \centering
    \includegraphics[width=\textwidth]{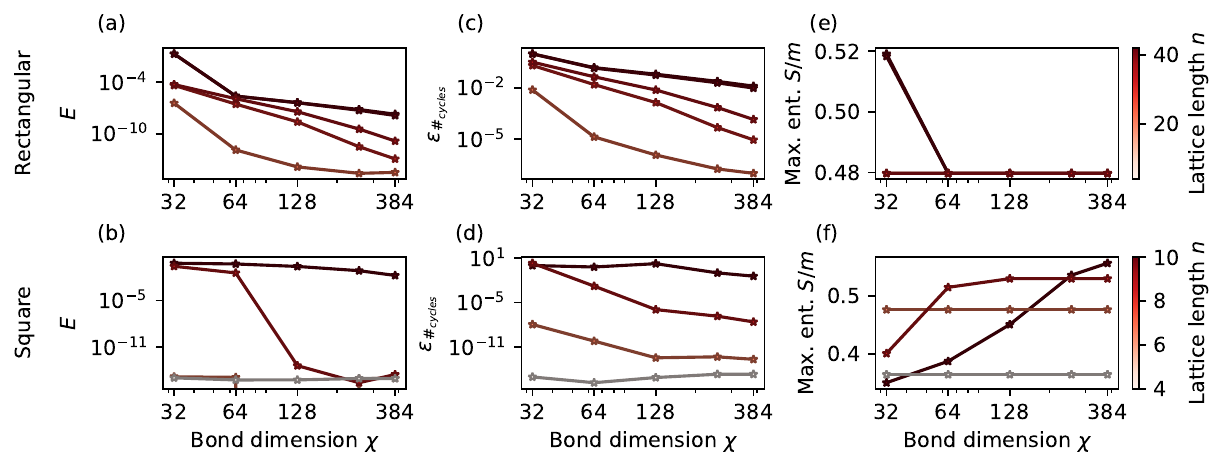}
    \caption{\textit{Convergence of the MPS approximation of $\ket{X_{\mathrm{HC}}}$ as a function of the bond dimension $\chi$.} 
    Panels~(a,c,e) correspond to rectangular lattices of size $6\times n$, while Panels~(b,d,f) correspond to square lattices of size $n\times n$. 
    \textbf{Panels~(a,b):} Tensor-network energy $E$. 
    \textbf{Panels~(c,d):} Relative error in the number of Hamiltonian cycles. 
    \textbf{Panels~(e,f):} Maximum bipartition entanglement entropy, normalized by the boundary area.}
    \label{fig:convergence_2}
\end{figure*}

In this appendix, we analyze the convergence of the various quality metrics and the entanglement of the MPS approximation discussed in the main text as a function of the bond dimension $\chi$, as shown in Fig.~\ref{fig:convergence_2}.

In Panels~(a--d), both the energy and the relative error in counting Hamiltonian cycles converge polynomially with the bond dimension, with the convergence rate decreasing as the system size increases. For square lattices, convergence only begins once the bond dimension exceeds a threshold value that grows rapidly with system size, preventing an efficient approximation.

This behaviour is clarified by examining the maximum bipartition entanglement entropy in Panels~(e) and (f). For rectangular lattices, a bond dimension of about $\chi \gtrsim 64$ already suffices to capture the relevant correlations. In contrast, for square lattices, the bond dimension required for entanglement convergence increases sharply with system size. Before convergence is reached, the entanglement is underestimated because the MPS truncation discards part of the Schmidt spectrum.

As anticipated in the main text, these observations are consistent with an area-law entanglement scaling.

\section{Quadratic speedup in counting}\label{app:amplification}

Here, we show that quantum amplitude amplification~\cite{Brassard,Grover_1998} quadratically speeds up the count of Hamiltonian cycles with respect to Monte Carlo sampling, provided a unitary algorithm
\begin{equation}
\hat {\mathcal{A}}\ket{0\dots0} = \ket{X_\text{HC}}
\end{equation}
that prepares the coherent quantum sample $\ket{X_\text{HC}}$. The algorithm $\hat{\mathcal{A}}$ could be, for example, a quantum annealing schedule ending with the Hamiltonian $\hat H_\text{HC}$.

Equation~\ref{eq:counting_exp_val} allows us to estimate the ratio $r$ between the total number of Hamiltonian cycles and the dimension of the lattice configuration space by measuring the expectation value of $
\bigotimes_{\substack{2\leq i\leq m\\2\leq j\leq n}}\hat P^x_{ij}$ for the state $\ket{X_\text{HC}}$. Let $\hat{\mathrm{H}}$ be the Hadamard gate. Since 
\begin{equation}
    \hat P^x =  \hat{\mathrm{H}} \hat P^+ \hat{\mathrm{H}}  =  \hat{\mathrm{H}} \ket{1}\bra{1} \hat{\mathrm{H}} \;,
\end{equation}
we can write
\begin{align}
r &= \bra{X_\text{HC}}\bigotimes_{\substack{2\leq i\leq m\\2\leq j\leq n}}\hat P^x_{ij}\ket{X_\text{HC}} \nonumber\\
&= \bra{X_\text{HC}}\hat{\mathrm{H}}^{\otimes n\cdot m}\left(\bigotimes_{\substack{2\leq i\leq m\\2\leq j\leq n}}\ket{1}\bra{1}_ {ij}\right)\hat{\mathrm{H}}^{\otimes n\cdot m}\ket{X_\text{HC}}\nonumber\\
&=\left|\bra{1\dots1}\left(\hat{\mathrm{H}}^{\otimes n\cdot m}\ket{X_\text{HC}}\right)\right|^2\nonumber\\
&=\left|\bra{1\dots1}\left(\hat{\mathrm{H}}^{\otimes n\cdot m}\hat{\mathcal{A}}\ket{0\dots0}\right)\right|^2\;.
\end{align}
Thus, $r$ is the probability of sampling the state $\ket{1\dots1}$ when the system is in the state $\hat{\mathrm{H}}^{\otimes n\cdot m}\ket{X_\text{HC}}$:

As anticipated in the main text, this probability is exponentially small in the lattice size. As a consequence, estimating $r$ up to a fixed acceptable relative error requires an exponential number of samples $N_S=\mathcal{O}(1/r)$. Despite this, Monte Carlo estimation remains a valid approach when considering squared lattices, where the state-of-art transfer-matrix counting methods also exhibit exponential cost. It is therefore meaningful to employ amplitude amplification to boost the estimation of $r$.

We apply amplitude amplification as follows. First, we prepare the system in the state $\hat{\mathrm{H}}^{\otimes n\cdot m}\ket{X_\text{HC}}$. 
Then, we repeatedly apply the operator
\begin{equation}
\hat W = \hat S_\text{HC} \hat S_1 \;,
\end{equation}
where
\begin{equation}
\hat S_1 = \id - 2\ket{1\dots 1}\bra{1\dots 1}
\end{equation}
reflects the system state with respect to the state $\ket{1\dots 1}$, and
\begin{equation}
\hat S_\text{HC} = \left(\id - 2\hat{\mathrm{H}}^{\otimes n\cdot m}\ket{X_\text{HC}}\bra{{X_\text{HC}}}\hat{\mathrm{H}}^{\otimes n\cdot m}\right)\;,
\end{equation}
reflects the system state with respect to $\hat{\mathrm{H}}^{\otimes n\cdot m}\ket{X_\text{HC}}$. We note that the knowledge of $\mathcal{\hat A}$ provides a recipe for implementing $\hat S_\text{HC}$. Indeed, we can write
\begin{align}
\hat S_\text{HC} &= \hat{\mathrm{H}}^{\otimes n\cdot m}\hat{\mathcal{A}}\left(\id - 2\ket{0\dots0}\bra{{0\dots 0}}\right)\hat{\mathcal{A}}^\dag \hat{\mathrm{H}}^{\otimes n\cdot m}\nonumber\\
&=\hat{\mathrm{H}}^{\otimes n\cdot m}\hat{\mathcal{A}}\hat{\mathrm{X}}^{\otimes n\cdot m}\hat S_1\hat{\mathrm{X}}^{\otimes n\cdot m}\hat{\mathcal{A}}^\dag \hat{\mathrm{H}}^{\otimes n\cdot m}\;,
\end{align}
where $\hat{\mathrm{X}}$ is a spin-flip gate. Finally, the operator $\hat S_1$ can be implemented efficiently on a quantum computer as a multi-controlled NOT gate.

After $\mathcal{O}(1/\sqrt{r})$ applications of the operator $\hat{W}$, the probability $p_{1\dots 1}$ of measuring $\ket{1\dots 1}$ approaches unity~\cite{Brassard,Grover_1998}. The number of samples needed to verify that $p_{1\dots 1}\approx 1$ does not depend on the specific starting state. Consequently, $N_C=\mathcal{O}(1/\sqrt{r})$ calls to $\hat {\mathcal{A}}$ are required to estimate $r$ with fixed relative accuracy, yielding a quadratic reduction compared to the $\mathcal{O}(1/r)$ samples needed by classical Monte Carlo methods.

\bibliography{main}

\end{document}